\documentclass[useAMS,usenatbib]{mnras}
\usepackage{widetext}
\usepackage{lscape}
\usepackage{graphicx}
\usepackage{dcolumn}
\usepackage{amsmath}
\usepackage{amsfonts}
\usepackage{amssymb}
\usepackage{dcolumn}% Align table columns on decimal point
\usepackage{enumerate}
\usepackage{color}
\usepackage{float}
\usepackage{subfig}
\usepackage{tabularx}
\usepackage{soul}
\usepackage{ulem}
\usepackage{bm}	

\newcolumntype{L}[1]{>{\raggedright\arraybackslash}p{#1}}
\newcolumntype{C}[1]{>{\centering\arraybackslash}p{#1}}
\newcolumntype{R}[1]{>{\raggedleft\arraybackslash}p{#1}}

\usepackage[utf8]{inputenc}

\begin{document}

\title[GEDE/PEDE models]{Generalized Emergent Dark Energy: observational Hubble data constraints and stability analysis}

\author[Hern\'andez-Almada, Leon, Maga\~na, Garc\'{\i}a-Aspeitia and Motta]{A. Hern\'andez-Almada$^{1}$\thanks{E-mail:ahalmada@uaq.mx}, Genly Leon$^{2}$\thanks{E-mail:genly.leon@ucn.cl}, Juan Maga\~na$^{3}$ \thanks{E-mail:jmagana@astro.puc.cl}, \newauthor Miguel A. Garc\'{\i}a-Aspeitia$^{4,5}$\thanks{E-mail:aspeitia@fisica.uaz.edu.mx}, V. Motta$^{6}$ \thanks{E-mail:veronica.motta@uv.cl}\\ $^{1}$ Facultad de Ingenier\'ia, Universidad Aut\'onoma de Quer\'etaro, Centro Universitario Cerro de las Campanas, 76010, \\ Santiago de Quer\'etaro, M\'exico\\
$^{2}$ Departamento  de  Matem\'aticas,  Universidad Cat\'olica del Norte, Avda.   Angamos  0610,  Casilla  1280  Antofagasta,  Chile \\
$^3$Instituto de Astrof\'isica \& Centro de Astro-Ingenier\'ia, Pontificia Universidad Cat\'olica de Chile, \\Av. Vicu\~na Mackenna, 4860, Santiago, Chile\\
$^{4}$ Unidad Acad\'emica de F\'isica, Universidad Aut\'onoma de Zacatecas, \\Calzada Solidaridad esquina con Paseo a la Bufa S/N C.P. 98060,Zacatecas, M\'exico.\\
$^{5}$Consejo Nacional de Ciencia y Tecnolog\'ia, Av. Insurgentes Sur 1582. \\ Colonia Cr\'edito Constructor, Del. Benito Ju\'arez C.P. 03940, Ciudad de M\'exico, M\'exico.\\
$^6$Instituto de F\'isica y Astronom\'ia, Universidad de Valpara\'iso, Avda. Gran Breta\~na 1111, Valpara\'iso, Chile.\\ }

\date{Accepted YYYYMMDD. Received YYYYMMDD; in original form YYYYMMDD}
\pubyear{2020}

%\label{firstpage}

\maketitle

\begin{abstract}
Recently, a phenomenologically emergent dark energy (PEDE) model was presented with a dark energy density evolving as $\widetilde{\Omega}_{\rm{DE}}(z)\,=\,\Omega_{\rm{DE,0}}\left[ 1 - {\rm{tanh}}\left( {\log}_{10}(1+z) \right) \right]$,  i.e. with no degree of freedom. Later on, a generalized model was
proposed by adding one degree of freedom to the PEDE model, encoded in the parameter $\Delta$. Motivated by these proposals,  we constrain the parameter space ($h,\Omega_m^{(0)}$) and ($h,\Omega_m^{(0)}, \Delta$) for PEDE and Generalized Emergent Dark Energy (GEDE) respectively, by employing the most recent observational (non-)homogeneous and differential age Hubble data. Additionally, we reconstruct the deceleration and jerk parameters and estimate yield values at $z=0$ of $q_0 = -0.784^{+0.028}_{-0.027}$ and $j_0 =  1.241^{+0.164}_{-0.149}$ for PEDE and $q_0 = -0.730^{+0.059}_{-0.067}$ and $j_0 =  1.293^{+0.194}_{-0.187}$ for GEDE using the homogeneous sample. We report values on the deceleration-acceleration transition redshift with those reported in the literature within $2\sigma$ CL. Furthermore, we perform a stability analysis of the PEDE and GEDE models to study the global evolution of the Universe around their critical points. Although the PEDE and GEDE dynamics are similar to the standard model, our stability analysis indicates that in both models there is an accelerated phase at early epochs of the Universe evolution.
\end{abstract}


%%%%%%%%%%%%%%%%%%%%%%%
\section{Introduction}
%%%%%%%%%%%%%%%%%%%%%%%

One of the most important challenges in modern cosmology is to elucidate the source of the accelerated expansion of the Universe, first evidenced by the high resolution observations of type Ia supernovae up to redshift $z\sim1.2$ \citep{Riess:1998,Perlmutter:1999} and then also confirmed by the acoustic peaks (position) of the cosmic microwave background radiation measurements \citep{Aghanim:2018}. In the framework of General Relativity (GR), the late cosmic acceleration is originated by an exotic component dubbed dark energy (DE). In the cosmological standard model, the nature of the dark energy is associated to the energy density of the vacuum ($\Lambda$), known as cosmological constant \citep{Zeldovich,Weinberg}. One of the main properties of $\Lambda$ is its equation of state, $w_{\Lambda}=-1$, implying an energy density constant over the cosmic time. The cosmological constant as DE has became a successful model to explain and fit several cosmological observations, however some theoretical aspects suggest that it might be necessary to consider a dynamical dark energy. For instance, there is no convincing fundamental hypothesis to explain the cosmological constant dominates the dynamics of the universe at late times, this is commonly known as the coincidence problem. Another crucial difficulty is to reconcile the estimations of the $\Lambda$ energy density from quantum field theory with those of the cosmological data. These problems have inspired plenty of models \citep[see][for a review]{Li:2011sd}
to explain that the late cosmic acceleration, some of them consider 
dynamical dark energy as a scalar field or a dark energy EoS parameterization \citep{Barboza:2008rh, ArmendarizPicon:2000ah,ArmendarizPicon:2000dh,Linder:2003,Chevallier:2000qy,Jassal:2004ej,sendra:2012, Wetterich:1988,Caldwell:1998,Caldwell:2002, Chiba:1998,Chiba:2000,Guo:2005,Magana:2017usz,Jaime:2019,Mario:2019}, interactions between dark energy and dark matter \citep{Cabral:2009, Bolotin:2015,DiValentino:2019ffd,Almada:2020}, viscous DE \citep{Cruz:2019wbl, Almada:2019},  but also models without dark energy where the Einstenian gravity is modified  \citep{Miguel:2018IJMPD,Miguel:2019a,Miguel:2018vbt,Miguel:2019b,Ovgun:2017iwg,Beto:2019EPJC}, and more recently, models that propose an emergent DE whose energy density is $\rho_{DE}(z) \propto \tanh(z)$ \citep{Mortonson_2009,dhawan:2020arx, PEDE:2019ApJ, PEDE:2020}. 
Reviews on Dark Energy (theory and observations) can be found in 
\citep[and references therein]{Capozziello:2005pa,Capozziello:2005tf,Copeland:2006wr,Tsujikawa:2010sc,Bamba:2012cp,Tsujikawa:2013fta}. 

On the other hand, studies with observational data pointed out that the DE could be evolving as function of the scale factor or redshift \citep{Holsclaw:2010PhRvD,Zhao:2017,Sola:2018sjf}. Recently, the evidence from two different groups (Planck and Supernova Projects) show that there is a significant tension in the value of $H_0$ (between $4.0\sigma$ and $5.8\sigma$), predicting, from Planck data a value of $H_0=67.4\pm0.5$km s$^{-1}$ Mpc$^{-1}$ \citep{Planck2018} with $1\%$ of precision, while for Supernovaes the value is $H_0=74.\pm1.4$km s$^{-1}$ Mpc$^{-1}$  (see \cite{Riess_2019}, and \cite{Verde:2019ivm} for details). Another interesting unexplained phenomena is related to an excess of radiation detected by the Experiment to Detect the Global Epoch of Reionization (EDGES, \citet{Bowman:2018yin} at $z\sim17$, which could be due to the interaction of dark matter with baryons but other explanations related to the presence of emergent DE in early epochs can not be discarded  \citep{Miguel:2019a,Miguel:2019b}.

Therefore, it is plausible to consider, as a natural extension to the cosmological constant, dynamical dark energy (DDE) models as the cause for the accelerated expansion of the Universe. Recently, \citet{PEDE:2019ApJ} introduced a phenomenological emergent dark energy model (PEDE), parameterizing the energy density of DE (with zero degree of freedom) using a hyperbolic tangent function, which is symmetric at logarithm scales (first attempts in the same line were done by \citet{Mortonson_2009}). In this model, the DE is negligible at early times but at late times the contribution of $\widetilde{\Omega}_{DE}$ increases, providing an alternative solution to the coincidence problem. The authors constrained the PEDE model using the latest data of Supernovae Ia (SNIa), Baryon Acoustic Oscillations (BAO) and the Planck measurements of Cosmic Microwave Background Radiation (CMB) and claim that it can solve the known tension problem with the Hubble constant. Later on, \citet{Pan:2019hac} constrained the PEDE model in a six parameter space using different observational data (mainly CMB), obtaining higher $H_{0}$ value than the standard model which reconcile the $H_0$ tension within $68\%$ of the confidence level. Moreover, the PEDE model is also studied by \citet{Koo:2020ssl} when they reconstruct the cosmic expansion from SNIa data using a non-parametric iterative smoothing method. They also show that PEDE SNIa constraints are consistent with those of the standard model. In this vein, \cite{PEDE:2020}, generalized the PEDE model, constructing the Generalized Emergent Dark Energy model (GEDE), which contains two new parameters: $\Delta$ indicates the model we are dealing (PEDE or $\Lambda$CDM) and a transition redshift $z_t$, with $\Omega_{DE}(z_{t})=\Omega_{m}(z_{t})$, establishing a relationship between the matter density parameter and $\Delta$ (i.e. not a free parameter).

Here, we revisit and constrain the free parameters of the PEDE and GEDE models using the latest compilation of observational Hubble data (OHD).
In addition, another vital study is the dynamical system analysis of these models. Dynamical systems analysis have provided to be very helpful to study the stability of several cosmological scenarios at background and perturbation levels \citep{Basilakos:2019dof}, for instance, Teleparallel Dark Energy \citep{Xu:2012jf,Karpathopoulos:2017arc,Cid:2017wtf}, Galileons \citep{Leon:2012mt,DeArcia:2015ztd,Giacomini:2017yuk,Dimakis:2017kwx,DeArcia:2018pjp}, Einstein-æther theories \citep{Latta:2016jix,Coley:2019tyx,Leon:2019jnu}, Hořava–Lifshitz theory \citep{Leon:2009rc,Leon:2019mbo}, Higher order Lagrangians \citep{Pulgar:2014cba}, non-linear electrodynamics \citep{Ovgun:2017iwg}, quintom models \citep{Lazkoz:2006pa,Lazkoz:2007mx,Leon:2018lnd}, modified Jordan-Brans-Dicke theory \citep{Cid:2015pja,Leon:2018skk,Giacomini:2020grc}, scalar field cosmologies \citep{Leon:2008de,Fadragas:2013ina,Fadragas:2014mra,Leon:2019iwj}, and other modified gravity models \citep{Leon:2013qh,Kofinas:2014aka,Leon:2014yua}. 
We investigate the stability of PEDE and GEDE models to search for different cosmic stages (i.e. radiation, matter, DE domination epochs) in order to demonstrate its feasibility with the standard $\Lambda$CDM model.

The paper is organized as follow: Sec. \ref{sec:CDF} we present the background cosmology of the PEDE and GEDE models. In Section \ref{sec:data}, we constrain the parameters of PEDE and GEDE models using the latest sample of OHD, discussing the results in Sec. \ref{sec:res}. Furthermore, in Sec. \ref{sec:stability} we discuss the stability of both models through a dynamical system analysis. Finally, we present our remarks and conclusions in Sec. \ref{sec:conclusions}.

%%%%%%%%%%%%%%%%%%%%%%%%%%%%%%%%%%%%%%%%%
\section{Phenomenological Emergent Dark Energy Cosmology} \label{sec:CDF}
%%%%%%%%%%%%%%%%%%%%%%%%%%%%%%%%%%%%%%%%%

In this section we introduced the phenomenological emergent dark energy model proposed by \citet{PEDE:2019ApJ} for which the DE is negligible at early times but it emerges at late times. We consider a flat Friedmann-Lemaitre-Robertson-Walker (FLRW) metric which contains matter (m, dark matter plus baryons), radiation (r), and PEDE. The dynamics of this Universe is described by the Friedmann equation and the continuity equation for each component as: 

\begin{subequations}
\label{non_min}
	\begin{eqnarray}
	&&H^2\equiv \left(\frac{\dot{a}}{a}\right)^2=\frac{8 \pi G}{3}(\rho_{DE}+\rho_{\rm{m}}+\rho_{r}),\label{Fried_rho}\\
	&&\dot{\rho}_{DE}+3H (1+ w_{DE})\rho_{DE}=0,\label{ce_de}\\
	&&\dot{\rho}_{\rm{m}}+3H(1+w_{m})\rho_{\rm{m}}=0,\label{ce_m}\\
	&&\dot{\rho}_{r}+3H(1+w_{r})\rho_{r}=0, \label{ce_rad}
	\end{eqnarray}
\end{subequations}
where $H$ is the Hubble parameter, $a$ the scale factor, $\rho_{i}$ is the energy density for each component, $w_{DE}=p_{DE}/\rho_{DE}$, $w_{m}=0$, $w_{r}=1/3$ are the equation of state for DE, matter and radiation respectively. By solving Eqs. \eqref{ce_de}, \eqref{ce_m}, \eqref{ce_rad} we can rewrite the Eq. \eqref{Fried_rho} in terms of the density parameters, $\Omega=\rho_{i}/\rho_{c}$\footnote{The critical density is defined as $\rho_{c}\equiv 3H^{2}/8\pi G$.}, and redshift, $z=1/(1+a)$, as 
\begin{equation}
    H(z)^{2}=H_{0}^{2}\left[\Omega_{m}^{(0)}\left(1+z\right)^{3}+\Omega_{r}^{(0)}\left(1+z\right)^{4}+ \widetilde{\Omega}_{\rm{DE}}(z) \right].
\end{equation}
The superscript $(0)$ denotes quantities evaluated at $z=0$ (or $a=1$) and $\widetilde{\Omega}_{\rm{DE}}(z)=\Omega_{\rm{DE}}^{(0)}f(z)$, where $\Omega_{\rm{DE}}^{(0)}=1-\Omega_{m}^{(0)}-\Omega_{r}^{(0)}$ from the flatness condition: 
\begin{equation}
    {\Omega}_{\rm{DE}}=1-\Omega_{m}-{\Omega}_{\rm{r}}. 
\label{eq:flatness}
\end{equation}
Notice that
\begin{equation}
f(z)\equiv \frac{\rho_{de}(z)}{\rho_{de}(0)}=
\mathrm{exp}\left(3\int^{z}_{0}\frac{1+w_{DE}(z)}{1+z}\mathrm{d}z\right).
\label{eq:fz}
\end{equation}
\citet{PEDE:2019ApJ} propose a phenomenological functional form for $f(z)$ and hence $\widetilde{\Omega}_{\rm{DE}}(z)$ as\footnote{Where it is defined  $\widetilde{\Omega}_{\rm{DE}}(z)\equiv \rho_{DE}/ \rho_{c}^{(0)}$.}
\begin{equation}
\widetilde{\Omega}_{\rm{DE}}(z)\,=\,\Omega_{\rm{DE}}^{(0)}\left[ 1 - {\rm{tanh}}\left( {\log}_{10}(1+z) \right) \right],
\label{eq:omegatilde_pede}
\end{equation}
\noindent
where $\widetilde{\Omega}_{\rm{DE}}\rightarrow 0$ at $z \rightarrow\infty$ and $\widetilde{\Omega}_{\rm{DE}}\rightarrow 1.4$ at $z \rightarrow -1$. Notice that

\begin{eqnarray}
    & \Omega_{DE}(z)= \frac{H_0^2}{H(z)^2}\widetilde{\Omega}_{\rm{DE}}(z)\nonumber\\
    & \,=\frac{H_0^2}{H(z)^2}\,\Omega_{\rm{DE}}^{(0)}\left[ 1 - {\rm{tanh}}\left( {\log}_{10}(1+z) \right) \right],  
\end{eqnarray}
Therefore, the dimensionless Friedmann equation results as
\begin{eqnarray}
    &&E(z)\equiv\frac{H(z)}{H_{0}}=\left\lbrace\right.\Omega_{m}^{(0)}\left(1+z\right)^{3}+\Omega_{r}^{(0)}\left(1+z\right)^{4}+\nonumber\\ &&\Omega_{\rm{DE}}^{(0)} \left[ 1 - {\rm{tanh}}\left( {\log}_{10}(1+z) \right) \right]\left. \right\rbrace^{1/2},\nonumber\\
\end{eqnarray}
where the radiation density parameter at current epoch is calculated as $\Omega_{r}^{(0)}=2.469 \times 10^{-5}h^{-2} (1+0.2271N_{eff})$, with $N_{eff}=3.04$ as the number of relativistic species \citep{Komatsu:2011}, and $h$ as the current Hubble dimensionless parameter.
The PEDE EoS can be calculated as
\begin{equation}
    w(z) \,=\,\frac{1}{3} \frac{d\,{\rm{ln}}\, \widetilde{\Omega}_{\rm{DE}}}{d z} (1+z)-1.
    \label{eq:eos_gen}
\end{equation}
By substituting \eqref{eq:omegatilde_pede} into Eq. \eqref{eq:eos_gen} results

\begin{equation}
    w(z)=\,-\frac{1}{3 {\rm{ln}}\, 10} \left({1+{\rm{tanh}}\left[{\log}_{10}\,(1+z)\right]}\right) -1. 
\end{equation}
The deceleration parameter $
q = -\ddot{a}a/\dot{a}^{2}$ can be rewritten in terms of redshift and $E(z)$ as:
\begin{eqnarray}
&&q(z)=\frac{(z+1)}{E(z)}\frac{dE(z)}{dz}-1,\nonumber\\
&&q(z)=\,-1+\frac{1}{2E(z)^2}\Big[3 \Omega_{m0} (z+1)^3+4 \Omega_{r0} (z+1)^4-\nonumber\\
&&\Omega_{\rm{DE}}^{(0)}\frac{\rm{sech}^2\left[\frac{\ln (z+1)}{\ln (10)}\right]}{\ln (10)}\Big].\nonumber\\
&&
\end{eqnarray}
For completeness we also calculate the jerk parameter, $j\equiv \dddot{a}/a\,H^{3}$
\begin{eqnarray}
&&j(z)=q(z)^2+\frac{(z+1)^2}{2E(z)^2}\frac{d^2E(z)^2}{dz^2}-\frac{(z+1)^2}{4E(z)^4}\nonumber\\&&\times\left(\frac{dE(z)^2}{dz}\right)^2, \label{j}
\end{eqnarray}
where
\begin{eqnarray}
&&\frac{dE(z)^2}{dz}=3\Omega_{m}^{(0)}\left(1+z\right)^{2}+4\Omega_{r}^{(0)}\left(1+z\right)^{3}-\nonumber\\ &&\Omega_{\rm{DE}}^{(0)}\frac{\rm{sech}^2\left[\frac{\ln (z+1)}{\ln (10)}\right]}{(1+z)\ln (10)} ,\\
&&\frac{d^2E(z)^2}{dz^2}=6\Omega_{m}^{(0)}\left(1+z\right)+12\Omega_{r}^{(0)}\left(1+z\right)^{2}+\nonumber\\&&\Omega_{\rm{DE}}^{(0)}\frac{\rm{sech}^2\left[\frac{\ln (z+1)}{\ln (10)}\right]}{(1+z)^2\ln (10)}+\Omega_{\rm{DE}}^{(0)}\frac{2\rm{sech}^2\left[\frac{\ln (z+1)}{\ln (10)}\right]}{(1+z)^2\ln ^2(10)}\times\nonumber\\&&\rm{tanh}\left[\frac{\ln (z+1)}{\ln (10)}\right],
\end{eqnarray}
which deviates from one, the jerk value for the cosmological constant.

%%%%%%%%%%%%%%%%%%%%%%%%%%%%%%%
\subsection{Generalized emergent dark energy} \label{Gen}
%%%%%%%%%%%%%%%%%%%%%%%%%%%%%%%

Recently, \citet{PEDE:2020} proposed a generalisation for the PEDE model also known as GEDE model by introducing
\begin{equation}
\widetilde{\Omega}_{\rm{DE}}(z)\,=\,\Omega_{\rm{DE}}^{(0)} \frac{ 1 - {\rm{tanh}}\left( {\Delta \log}_{10}(\frac{1+z}{1+z_{t}}) \right)}{ 1 + {\rm{tanh}}\left(\Delta {\log}_{10}(1+z_{t}) \right)},
\label{eq:omega_pede}
\end{equation}
where $z_t$ is a transition redshift, $\Omega_{DE}(z_t)=\Omega^{(0)}_{m}(1+z_t)^3$, $\Delta$ is an appropriate dimensionless non-negative free parameter with the characteristic that if $\Delta=0$ the $\Lambda$CDM model is recovered, and when $\Delta=1$ and $z_t=0$ the previously PEDE model is recovered. As $z_{t}$ can be related to $\Omega_{m}^{(0)}$ and $\Delta$, then $z_{t}$ is not a free parameter. Notice that the DE density parameter is given by
\begin{eqnarray}
&{\Omega}_{\rm{DE}}\, =\frac{H_0^2}{H^2} (1-\Omega_{m}^{(0)}-\Omega_{r}^{(0)})\frac{ 1 - {\rm{tanh}}\left( {\Delta \log}_{10}(\frac{1+z}{1+z_{t}}) \right)}{ 1 + {\rm{tanh}}\left(\Delta {\log}_{10}(1+z_{t}) \right)}.
\label{eq:omegade_gene}
\end{eqnarray}

The GEDE Friedmann equation is given by
\begin{eqnarray}
    &&E(z)\equiv\frac{H(z)}{H_{0}}=\Big[\Omega_{m}^{(0)}\left(1+z\right)^{3}+\Omega_{r}^{(0)}\left(1+z\right)^{4}+\nonumber\\ &&\Omega_{\rm{DE}}^{(0)} \frac{ 1 - {\rm{tanh}}\left( {\Delta\, \log}_{10}(\frac{1+z}{1+z_{t}}) \right)}{ 1 + {\rm{tanh}}\left(\Delta \,{\log}_{10}(1+z_{t}) \right)} \Big]^{1/2}.\nonumber\\
\end{eqnarray}
The EoS for GEDE model is given by
\begin{equation}
\label{wPEDE}
    w(z)=\,-\frac{\Delta}{3 {\rm{ln}}\, 10}\left({1+{\rm{tanh}}\left[\Delta{\log}_{10}\,\frac{(1+z)}{1+z_{t}}\right]}\right) -1. 
\end{equation}
The deceleration parameter reads
\begin{eqnarray}
&&q(z)=-1+\frac{1}{2E(z)^2}\Big[3\Omega_{m}^{(0)}\left(1+z\right)^{3}+4\Omega_{r}^{(0)}\left(1+z\right)^{4}-\nonumber\\&&\Omega_{\rm{DE}}^{(0)}\frac{\Delta}{\ln(10)}\frac{\rm{sech}^2\left[\frac{\Delta\ln\left(\frac{1+z}{1+z_t}\right)}{\ln(10)}\right]}{1+\rm{tanh}(\Delta\log_{10}(1+z_t))}\Big].
\end{eqnarray}
As a complement, we also calculate the GEDE jerk parameter using Eq. \eqref{j}, where
\begin{eqnarray}
&&\frac{dE(z)^2}{dz}=3\Omega_{m}^{(0)}\left(1+z\right)^2+4\Omega_{r}^{(0)}\left(1+z\right)^3-\nonumber\\&&\Omega_{\rm{DE}}^{(0)}\frac{\Delta}{\ln(10)(1+z)}\frac{\rm{sech}^2\left[\frac{\Delta\ln\left(\frac{1+z}{1+z_t}\right)}{\ln(10)}\right]}{1+\rm{tanh}(\Delta\log_{10}(1+z_t))}, \\
&&\frac{d^2E(z)^2}{dz^2}=6\Omega_{m}^{(0)}\left(1+z\right)+12\Omega_{r}^{(0)}\left(1+z\right)^2+\nonumber\\&&\Omega_{\rm{DE}}^{(0)}\frac{\Delta}{\ln(10)(1+z)^2}\frac{\rm{sech}^2\left[\frac{\Delta\ln\left(\frac{1+z}{1+z_t}\right)}{\ln(10)}\right]}{1+\rm{tanh}(\Delta\log_{10}(1+z_t))}+\nonumber\\&&\Omega_{\rm{DE}}^{(0)}\frac{2\Delta^2}{\ln^2(10)(1+z)^2}\frac{\rm{sech}^2\left[\frac{\Delta\ln\left(\frac{1+z}{1+z_t}\right)}{\ln(10)}\right]}{1+\rm{tanh}(\Delta\log_{10}(1+z_t))}\times\nonumber\\&&\rm{tanh}\left[\frac{\Delta\ln\left(\frac{1+z}{1+z_t}\right)}{\ln(10)}\right].
\end{eqnarray}

%%%%%%%%%%%%%%%%%%%%%%%%%%%%%%%%%%%%%%%%
\section{Observational constraints} \label{sec:data}
%%%%%%%%%%%%%%%%%%%%%%%%%%%%%%%%%%%%%%%%%

A canonical test is to confront a cosmological model with the observational Hubble data (OHD) which gives direct measurement of expansion rate of the Universe. Currently, the OHD sample is obtained from the differential age technique \citep[DA,][]{Jimenez:2002ApJ,Moresco:2012JCAP} and BAO measurements. In this work, we consider the sample compiled by \cite{Magana:2018}, which consists of $51$ points in the redshift region $0.07<z<2.36$. It is worth to note that $31$ data points come from the cosmic chronometers, i.e. passive galaxies, using the DA technique which are cosmological-model-independent. However, $20$ data points of this sample are estimated from BAO measurements under different fiducial cosmologies (based on $\Lambda$CDM), which could provide biased constraints. Nevertheless, \citet{Magana:2018} present also homogeneous BAO OHD points calculated using the sound horizon at the drag epoch from Planck measurements. Here, we use the full sample with non-homogeneous and homogeneous OHD data points from BAO, and OHD from DA method. Thus, the figure-of-merit is given by
\begin{equation}\label{eq:chi2_ohd}
\chi^2_{OHD} = \sum_{i=1}^{N_{i}} \left( \frac{H_{th}(z_i, {\bf \Theta}) - H_{obs}(z_i)}{\sigma_{obs}^i} \right)^2 \,,
\end{equation}
where $N_{i}$ is the number of data points, $H_{th}(z_i, {\bf \Theta})-H_{obs}(z_i)$ denotes the difference between the theoretical Hubble parameter with parameter space ${\bf \Theta}=(h,\Omega_{dm}^{(0)})$ and $(h,\Omega_{dm}^{(0)}, \Delta)$ for PEDE and GEDE models respectively, and the observational one at the redshift $z_i$, and $\sigma_{obs}^i$ is the uncertainty of $H_{obs}^i$.

To constrain the PEDE and GEDE cosmological parameters we perform
a Markov chain Monte Carlo (MCMC) analysis employing the emcee Python
module \citep{Emcee:2013}. We consider Gaussian likelihoods $\mathcal{L}\propto e^{-\chi^{2}/2}$, a Gaussian prior over $h$ centered at $h=0.7403\pm 0.0142$ \citep[][R19, hereafter]{Riess_2019} and a flat prior over $\Omega_{m}^{(0)} : [0,1]$ for both, PEDE and GEDE models. Additionally, we consider a flat prior on $\Delta: [0,10]$. Notice that the parameter $z_{t}$ presented in the GEDE model is related to the parameter $\Delta$ through the condition $\Omega_{DE}(z_t)=\Omega_{m}(z_t)$. As a complement, we perform a similar analysis but alternatively using a flat prior on $h:[0,1]$.  Our analysis consider a burn-in phase which is stopped when the Gelman-Rubin convergence criteria ($<1.1$) is fulfilled and a MCMC phase with $3000$ steps and $500$ walkers for each one.

%%%%%%%%%%%%%%%%%%%%%%%%%%%%%%%%
\subsection{Results}\label{sec:res}
\begin{figure}
\centering
\includegraphics[width=7cm,scale=0.5]{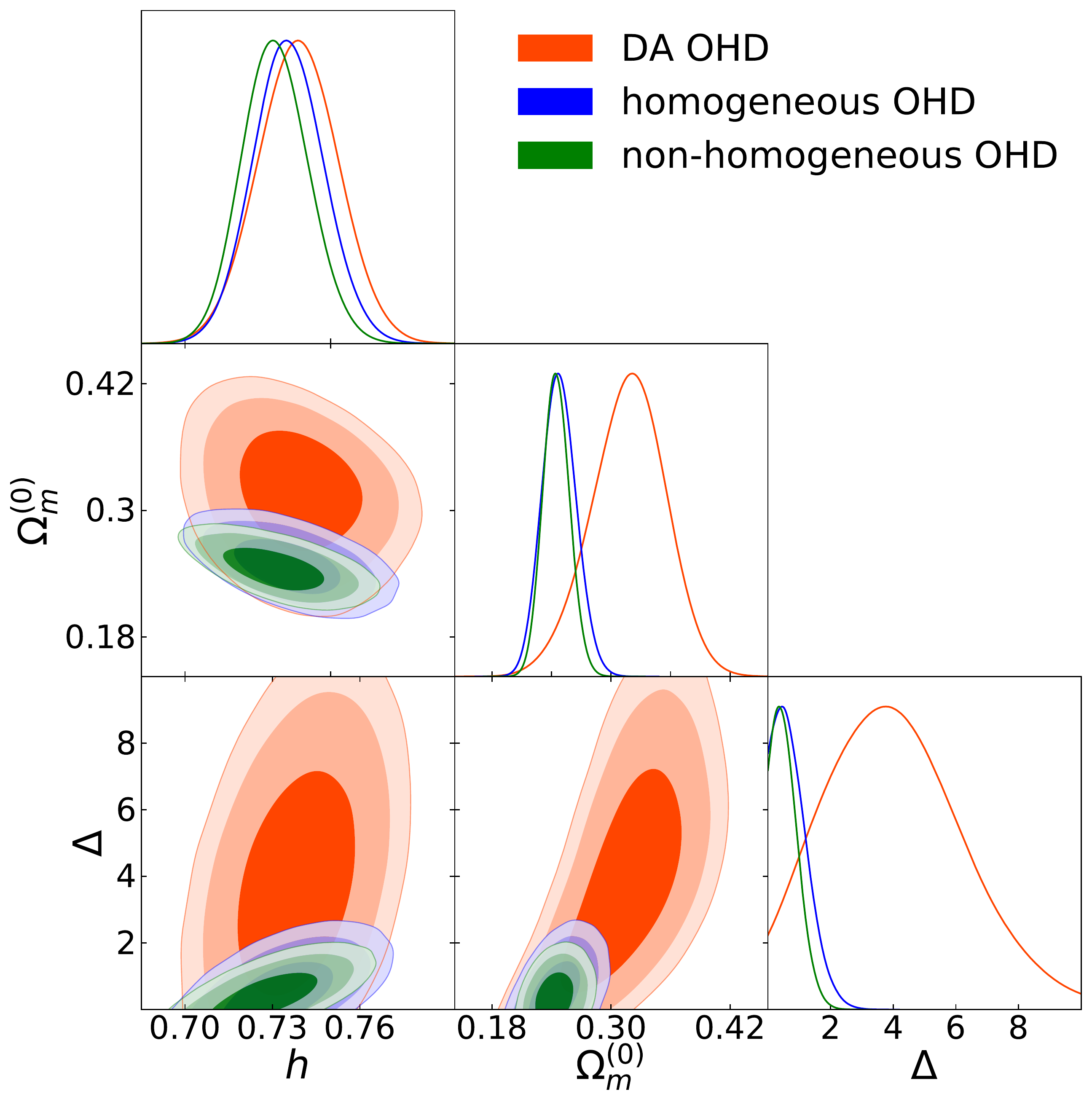}\\ 
\includegraphics[width=7cm,scale=0.5]{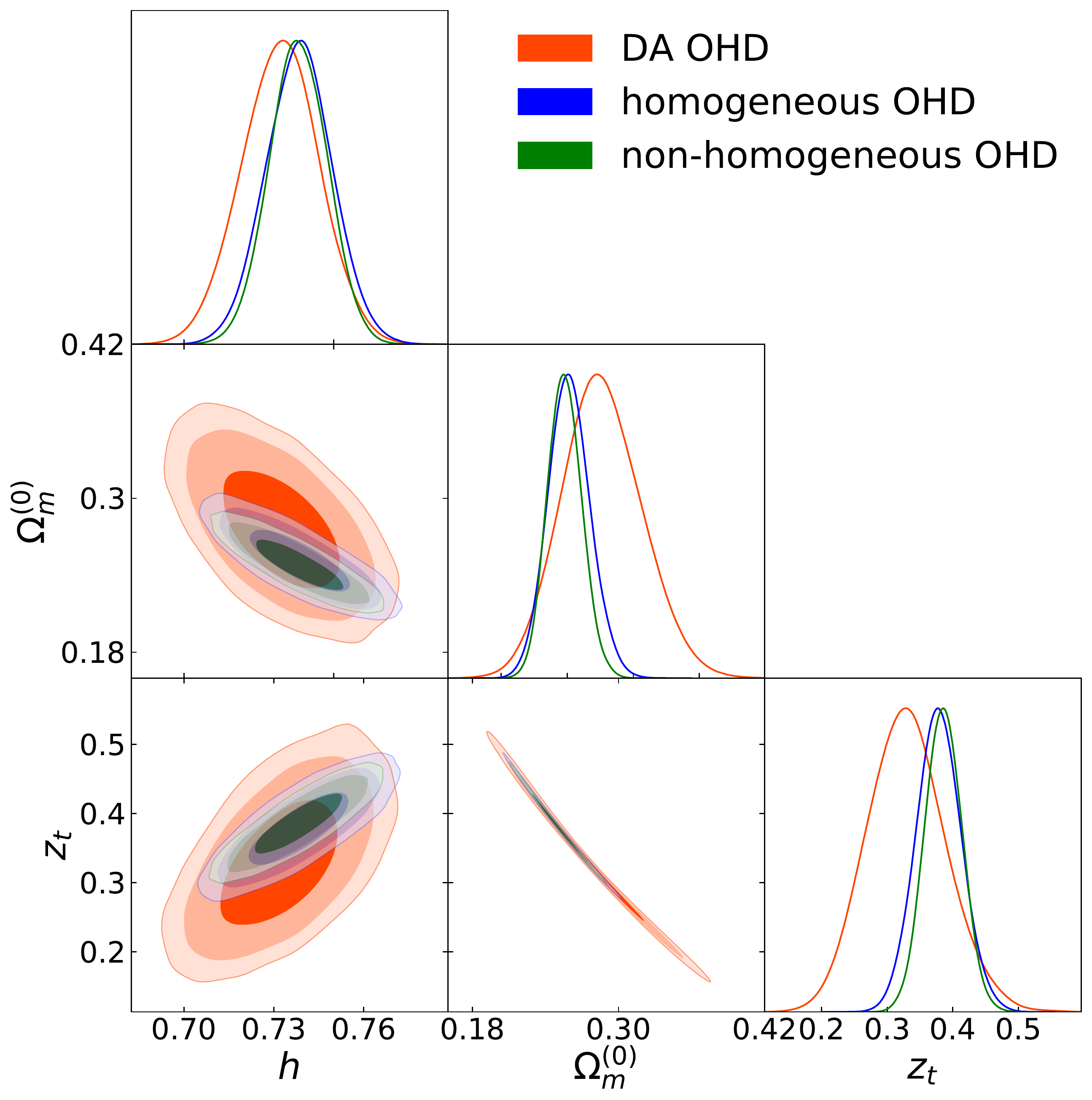}\\
\includegraphics[width=7cm,scale=0.5]{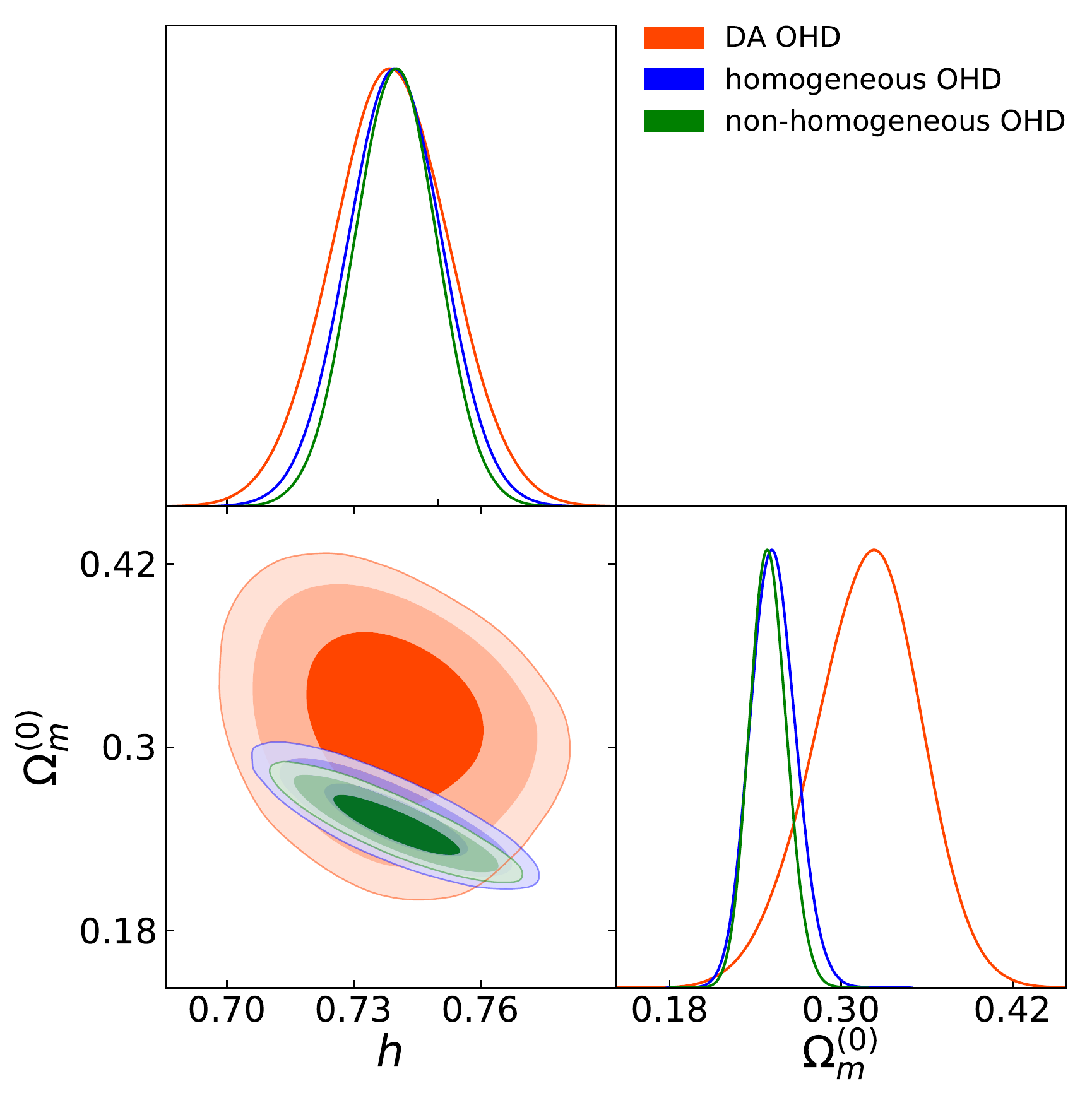}
\caption{1D posterior distributions and 2D contours of the free parameters for GEDE (top panel) and PEDE with the constraint $\Omega_m(z_t)= \Omega_{de}(z_t)$ (middle panel) and the case $z_t=0$ (bottom panel) models at $1\sigma$, $2\sigma$, $3\sigma$ CL (from darker to lighter respectively). The orange, blue and green contours correspond to the space constrained using DA and (non-) homogeneous OHD respectively.} 
\label{fig:contours:PEDE}
\end{figure}

\begin{figure*}
\centering
\includegraphics[width=5.5cm,scale=0.31]{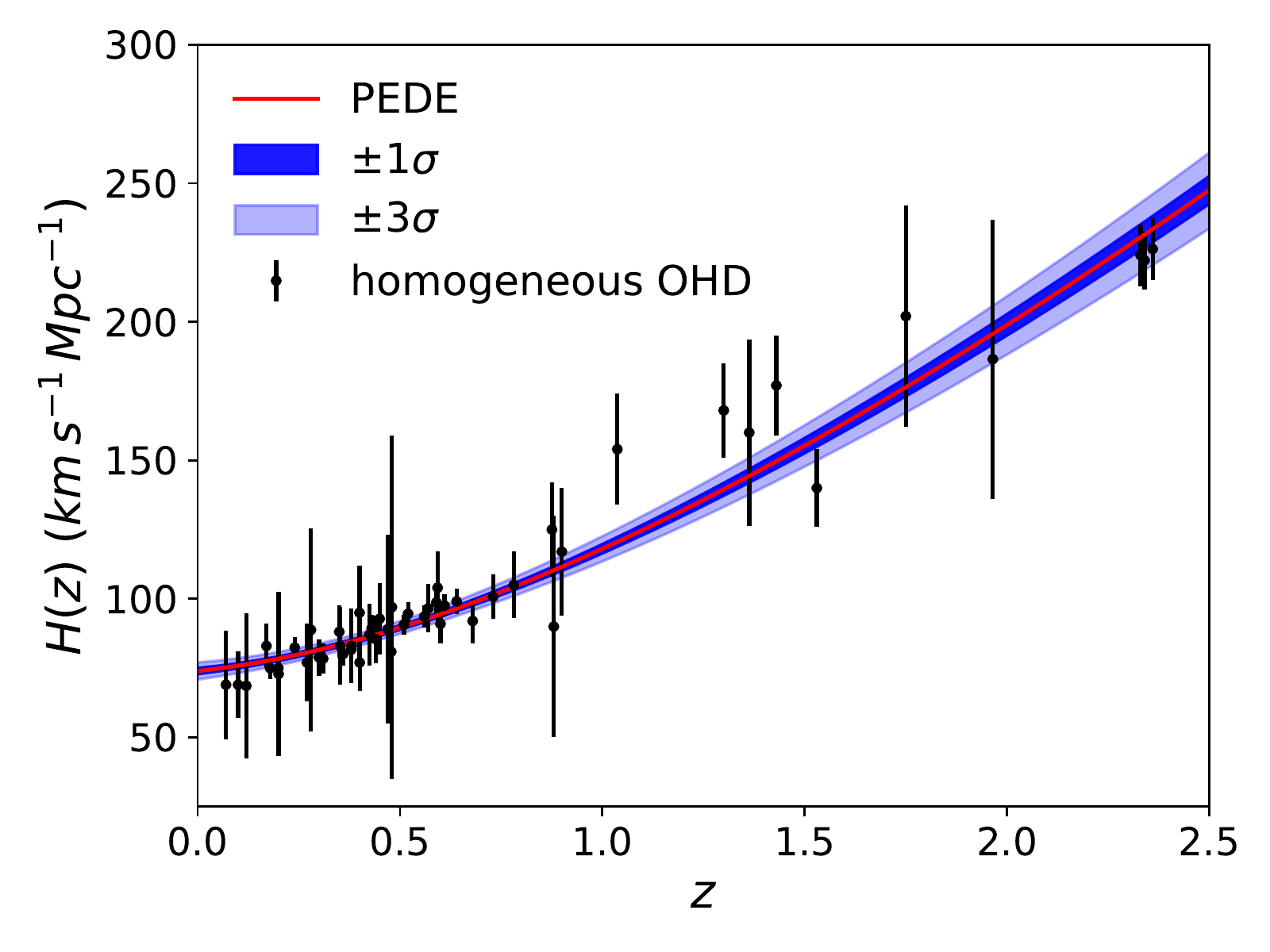} 
\includegraphics[width=5.5cm,scale=0.31]{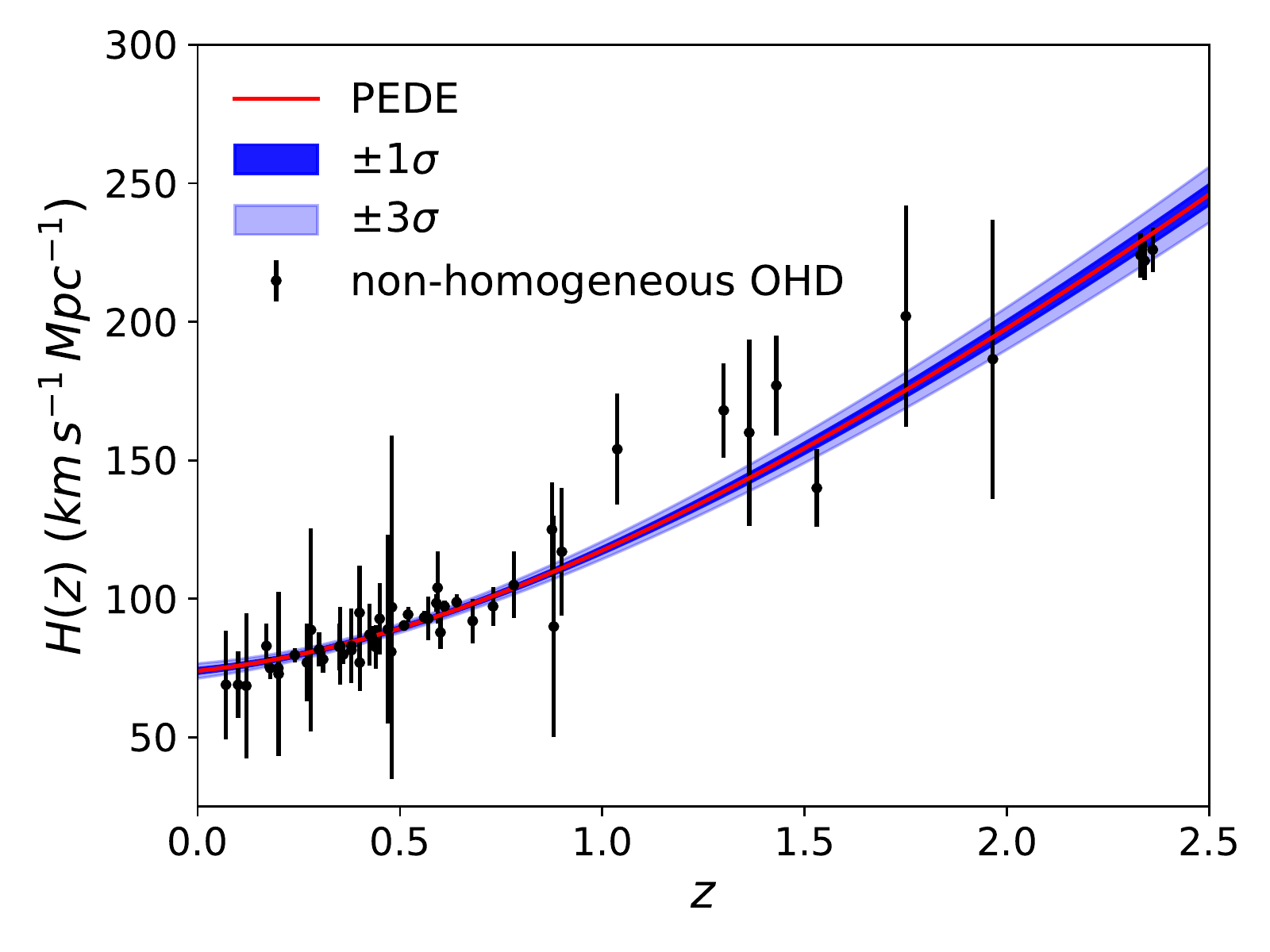} 
\includegraphics[width=5.5cm,scale=0.31]{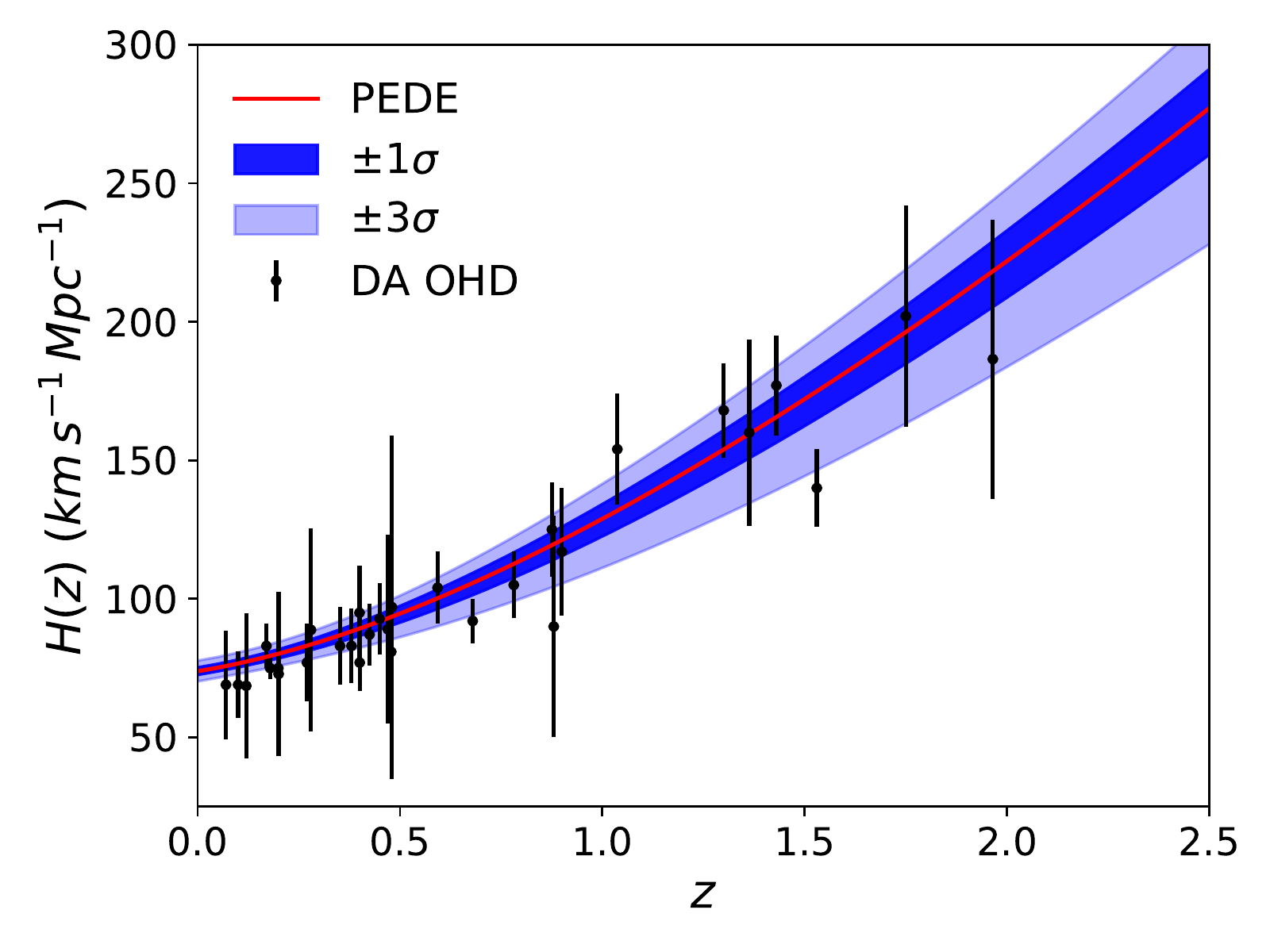} \\
\includegraphics[width=5.5cm,scale=0.31]{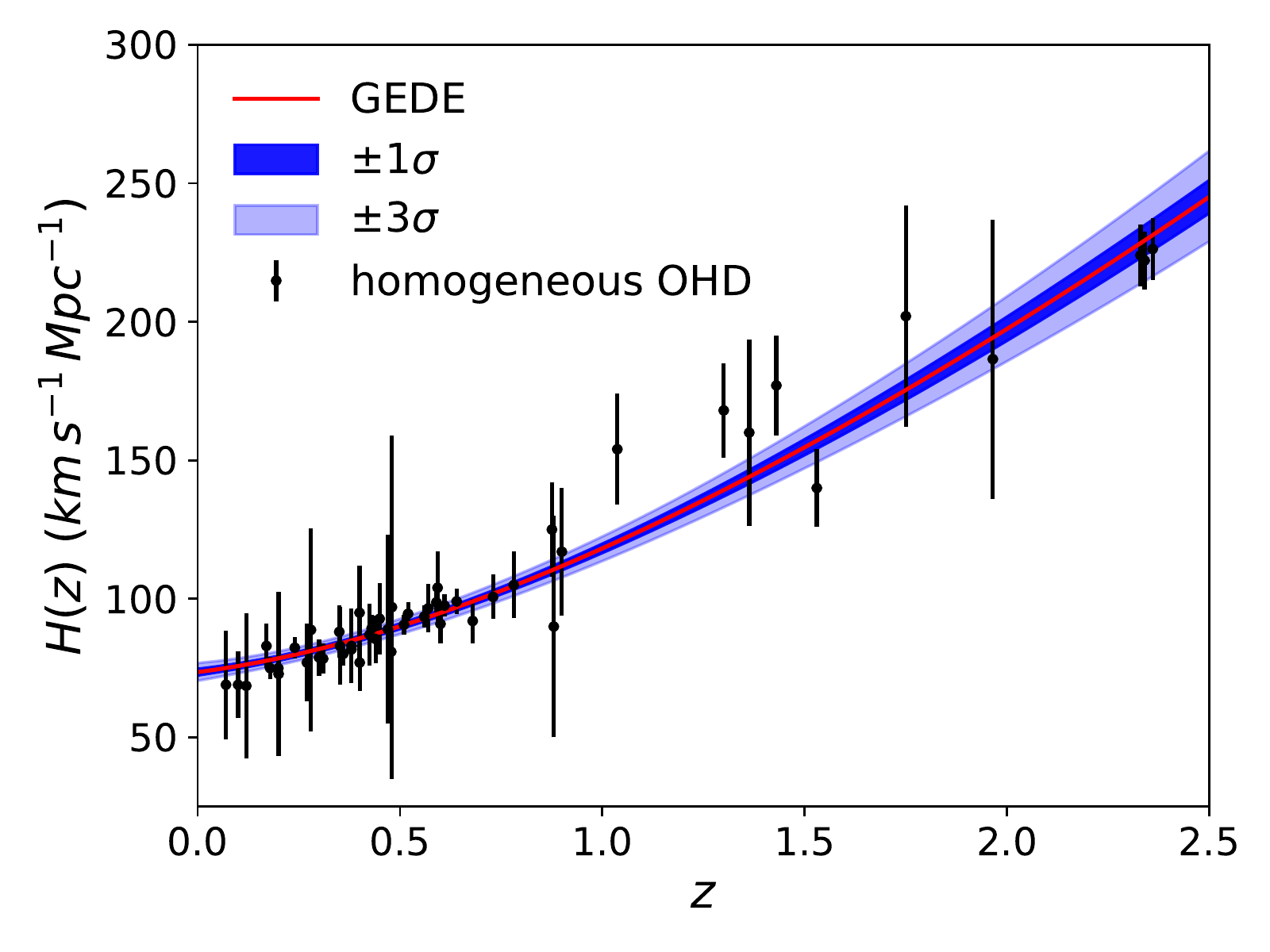} 
\includegraphics[width=5.5cm,scale=0.31]{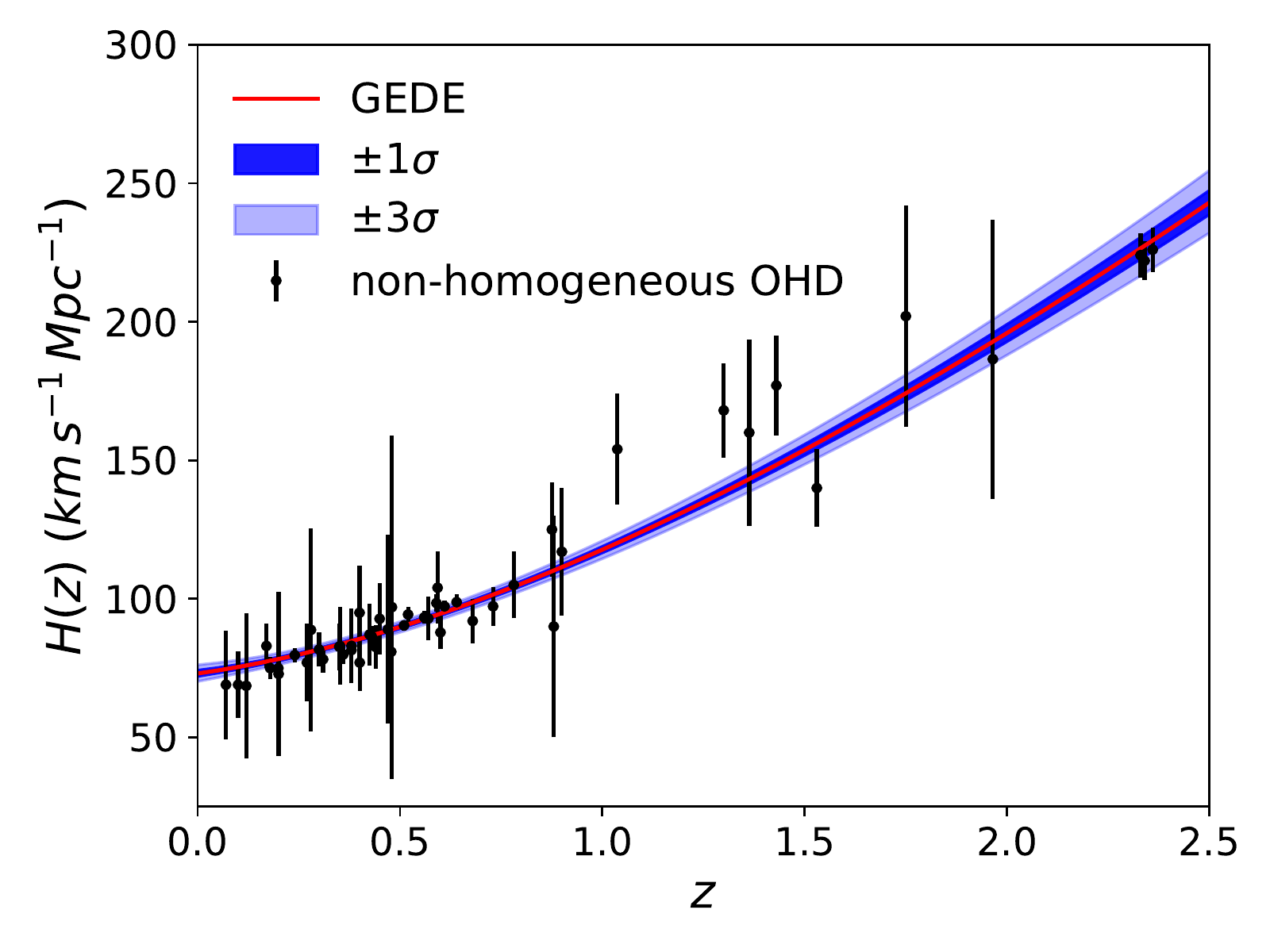}
\includegraphics[width=5.5cm,scale=0.31]{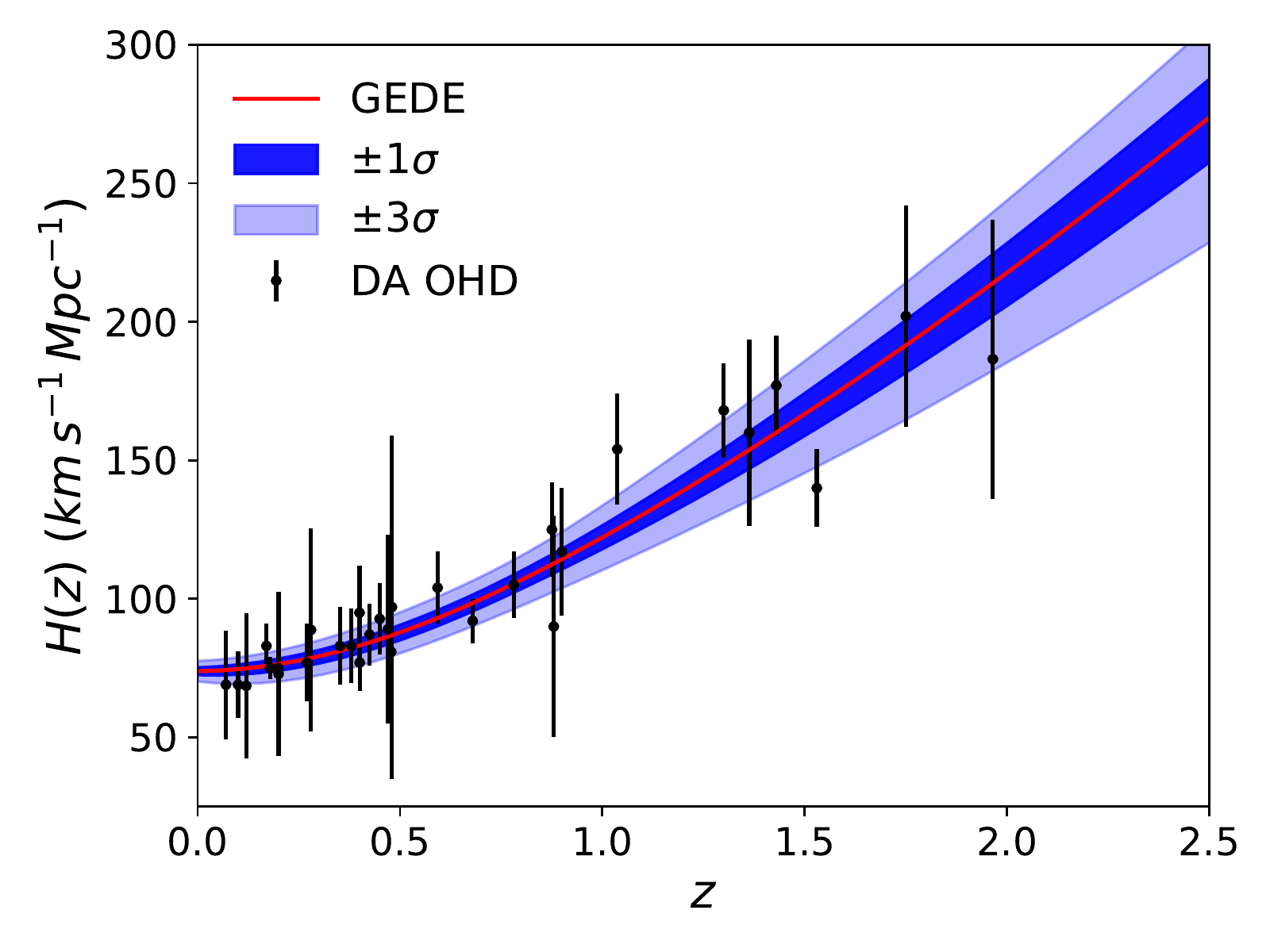} \\
\caption{Best fits over (non-)homogeneous and DA OHD sample at left, middle and right side of the panel for PEDE (top panel) and GEDE (bottom panel). The darker (lighter) band represents the uncertainty at $1\sigma$ ($3\sigma$) CL.}
\label{fig:Hz}
\end{figure*}

%%%%%%%%%%%%%%%%%%%%%%%%%%%%%%%%
In this section we report our results obtained in Bayesian analysis. In Table \ref{tab:bestfits} are provided the mean values for the parameters and their uncertainties estimated at $1\sigma$ in both scenarios and using the homogeneous, non-homogeneous and DA OHD. Additionally, we also report the parameter mean values when a flat prior over $h$ is considered. These are in agreement with those obtained using a Gaussian prior on $h$. Our constraints are very similar to those obtained by \citet{PEDE:2019ApJ}, estimating a deviation on $\Delta$ within $1\sigma$ CL with the one estimated by \citet{PEDE:2019ApJ} from a CMB+ $h$ (R19) joint analysis. Figure \ref{fig:contours:PEDE} shows the 2D confidence region at 68\% ($1\sigma$), 95\% ($2\sigma$) and 99.7\% ($3\sigma$) of the free parameters for GEDE (top panel) and PEDE (bottom panel) models, using the homogeneous, non-homogeneous and DA OHD, respectively. Moreover, their 1D posterior distributions are presented. 
Regarding the generalization of PEDE discussed by \citet{PEDE:2019ApJ}, consisting in the addition of the parameter $z_t$, we found that our constraints on the space ($h$, $\Omega_m^{(0)}$) presented in Fig. \ref{fig:contours:PEDE} are independent of the selected value for $z_t$ (see Appendix \ref{sec:AppendixA}). Nevertheless, we have also constrained $z_t$ by requiring the condition $\Omega_{m}(z_t)=\Omega_{DE}(z_t)$. We found mean values of $z_t=0.378, 0.386$, and $0.328$ for homogeneous, non-homogeneous, and DA data, respectively ($h$ Gaussian prior). Results are shown in Table \ref{tab:bestfits} and in the middle panel of Fig. \ref{fig:contours:PEDE}, which represents the corresponding constrained space ($h$, $\Omega_m^{(0)}, z_t$). Notice that there is no significant differences on the $h$ and $\Omega_{m}$ bounds considering $z_t=0$ and  $z_{t}$ constrained to $\Omega_{m}(z_t)=\Omega_{DE}(z_t)$.
It is worth to note that although the homogeneous sample provides slightly
broader confidence contours than those obtained with the non-homogeneous sample, the constraints are less (cosmology-model) unbiased.  Regarding DA OHD constraints, although we obtain the less restricted regions of the model parameters, they are completely unbiased (model independent). As it is expected, we find an anti-correlation relation between $\Omega_{\rm{m}}^{(0)}$ and $h$ for both models. For GEDE model, we also observe a positive correlation between $\Delta$ and $h$.
For the GEDE model, our $\Delta$ constraints are in tension with $\Delta=1.13\pm 0.28$ obtained by \citet{PEDE:2020} employing CMB and the $H_{0}$ measurements. Additionally, in the case of LCDM, it is interesting to observe that the best-fit value of $\Delta$ using DA OHD has the largest deviation compared to those obtained with the homogeneous or non-homogeneous samples. Figure \ref{fig:Hz} shows the comparison of the Hubble parameter in GEDE and PEDE cosmologies with the observational ones including non-homogeneous, homogeneous and DA OHD points. Notice that both models provide a good fit to the data. In addition, our estimates on $H_0$ and $\Omega_m^{(0)}$ are consistent within $1.2\sigma$ of those values obtained by \cite{Pan:2019hac} and \cite{Riess_2019}, alleviating the tension with the results obtained by Planck satellite.

\begin{table*}
\caption{Mean values of the free parameters for GEDE and PEDE models using homogeneous, non-homogeneous and DA OHD and a Gaussian prior on $h=0.7403\pm 0.0142$ \citep{Riess_2019}. The last column shows the estimated redsfhit $z_t$ using the condition $\Omega_m(z_t) = \Omega_{DE}(z_t)$. The uncertainties reported correspond to $1\sigma$ confidence level. In parenthesis are the best fit
values when a flat prior on $h$ is considered in the region $[0,1]$.}
\centering
\resizebox{\textwidth}{!}{%
\begin{tabular}{|lccccc|}
\hline
Sample     &    $\chi^2$     &  $h$ & $\Omega_m^{(0)}$ & $\Delta$ & $z_t$ \\
\hline
\multicolumn{6}{|c|}{PEDE} \\ [0.9ex]
homogeneous OHD  & $24.5$ ($24.5$) & $0.740^{+0.011}_{-0.011}$ ($0.738^{+0.018}_{-0.018}$)& $0.252^{+0.016}_{-0.015}$ ($0.254^{+0.024}_{-0.022}$) & $1.0$  & $0$ \\ [0.9ex]
non-homogeneous OHD    & $32.1$ ($32.1$) & $0.740^{+0.010}_{-0.010}$ ($0.740^{+0.014}_{-0.014}$) & $0.249^{+0.013}_{-0.013}$ ($0.249^{+0.018}_{-0.016}$) & $1.0$  & $0$\\ [0.9ex]
DA OHD    & $14.7$ ($14.6$) & $0.739^{+0.014}_{-0.014}$ ($0.723^{+0.049}_{-0.044}$) & $0.319^{+0.035}_{-0.039}$ ($0.329^{+0.057}_{-0.045}$) & $1.0$  & $0$\\ [0.9ex]
homogeneous OHD  & $24.2$ ($24.2$) & $0.739^{+0.011}_{-0.011}$ ($0.735^{+0.018}_{-0.018}$) & $0.251^{+0.016}_{-0.015}$ ($0.255^{+0.024}_{-0.022}$) & $1.0$  & $0.378^{+0.035}_{-0.034}$ ($0.371^{+0.049}_{-0.049}$) \\ [0.9ex]
non-homogeneous OHD    & $31.6$ ($31.6$) & $0.738^{+0.010}_{-0.010}$ ($0.736^{+0.013}_{-0.013}$)& $0.248^{+0.013}_{-0.013}$ ($0.250^{+0.017}_{-0.016}$) & $1.0$  & $0.386^{+0.028}_{-0.028}$ ($0.381^{+0.037}_{-0.037}$)\\ [0.9ex]
DA OHD    & $16.1$ ($14.4$)& $0.732^{+0.013}_{-0.013}$ ($0.691^{+0.032}_{-0.032}$) & $0.275^{+0.031}_{-0.029}$ ($0.333^{+0.064}_{-0.054}$)& $1.0$  & $0.328^{+0.060}_{-0.058}$ ($0.226^{+0.096}_{-0.096}$)\\ [0.9ex]

\hline
\multicolumn{6}{|c|}{GEDE} \\ [0.9ex]
homogeneous OHD  & $23.7$ ($23.0$) & $0.735^{+0.012}_{-0.012}$ ($0.725^{+0.023}_{-0.020}$) & $0.247^{+0.018}_{-0.017}$ ($0.256^{+0.025}_{-0.022}$) & $0.690^{+0.624}_{-0.457}$ ($0.533^{+0.712}_{-0.390}$) & $0.403^{+0.058}_{-0.057}$ ($0.385^{+0.058}_{-0.056}$) \\ [0.9ex]
non-homogeneous OHD    & $30.2$ ($28.6$)& $0.731^{+0.012}_{-0.011}$ ($0.718^{+0.017}_{-0.015}$) & $0.245^{+0.014}_{-0.013}$ ($0.255^{+0.018}_{-0.017}$) & $0.539^{+0.470}_{-0.352}$ ($0.332^{+0.472}_{-0.244}$) & $0.417^{+0.044}_{-0.043}$ ($0.403^{+0.043}_{-0.043}$)\\ [0.9ex]
DA OHD  & $14.7$ ($14.6$) & $0.739^{+0.014}_{-0.014}$ ($0.723^{+0.048}_{-0.044}$) & $0.319^{+0.036}_{-0.039}$ ($0.329^{+0.057}_{-0.046}$) & $3.930^{+2.304}_{-2.083}$ ($3.264^{+3.258}_{-2.230}$) & $0.183^{+0.094}_{-0.057}$ ($0.174^{+0.083}_{-0.064}$) \\ [0.9ex]

\hline
\end{tabular}}
\label{tab:bestfits}
\end{table*}


\begin{figure*}
\centering
\includegraphics[width=5.5cm,scale=0.55]{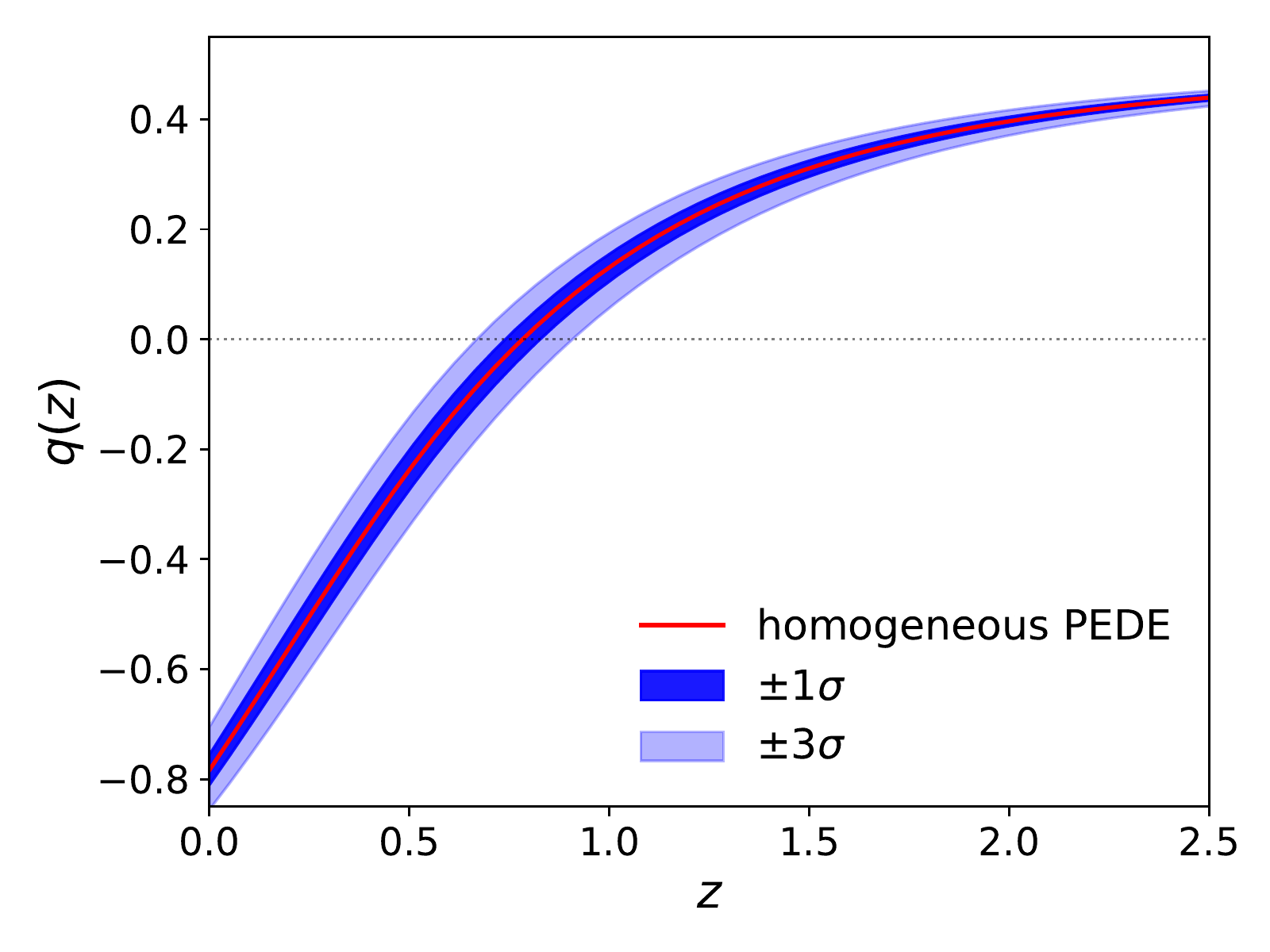} 
\includegraphics[width=5.5cm,scale=0.55]{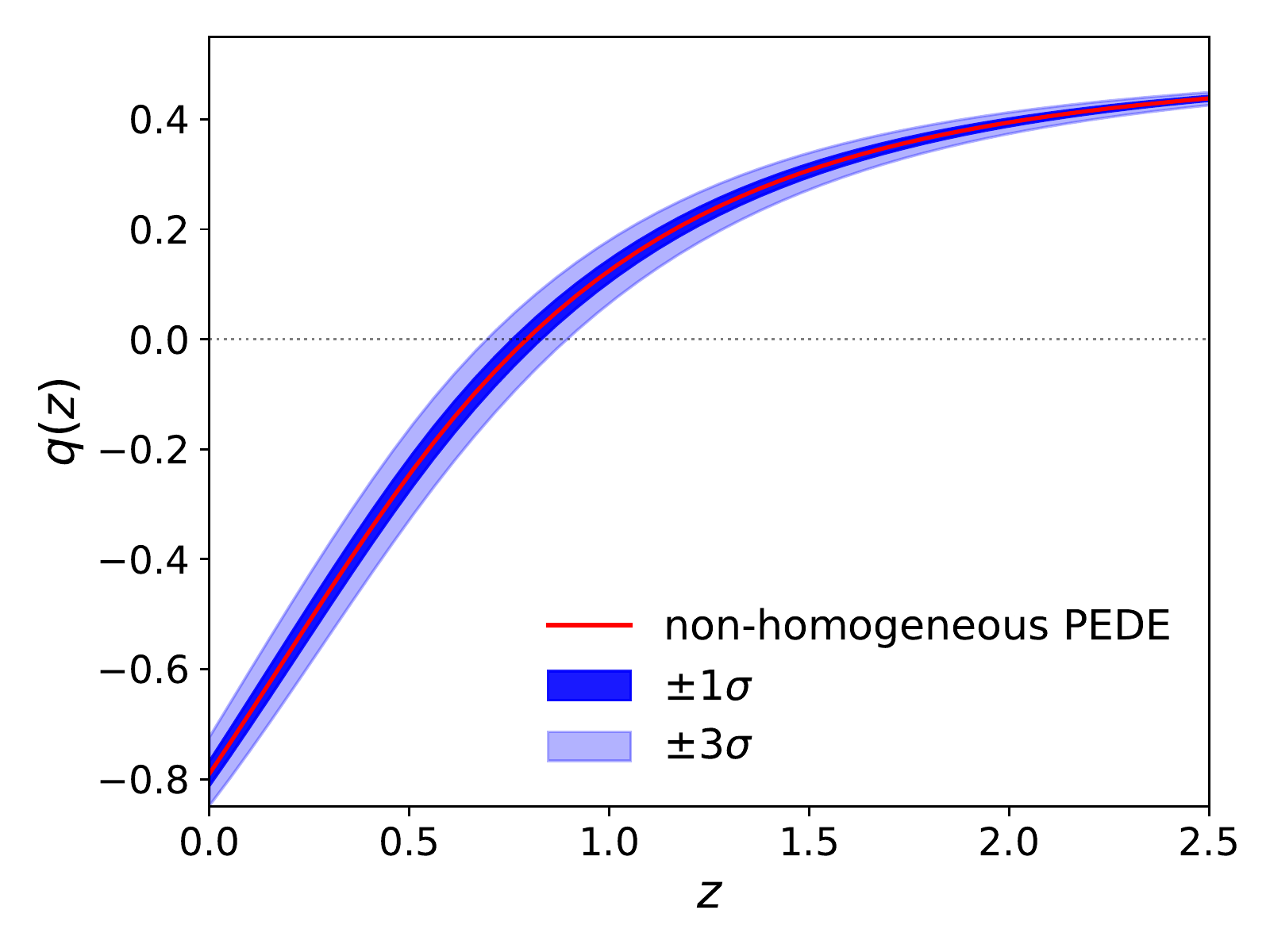} 
\includegraphics[width=5.5cm,scale=0.55]{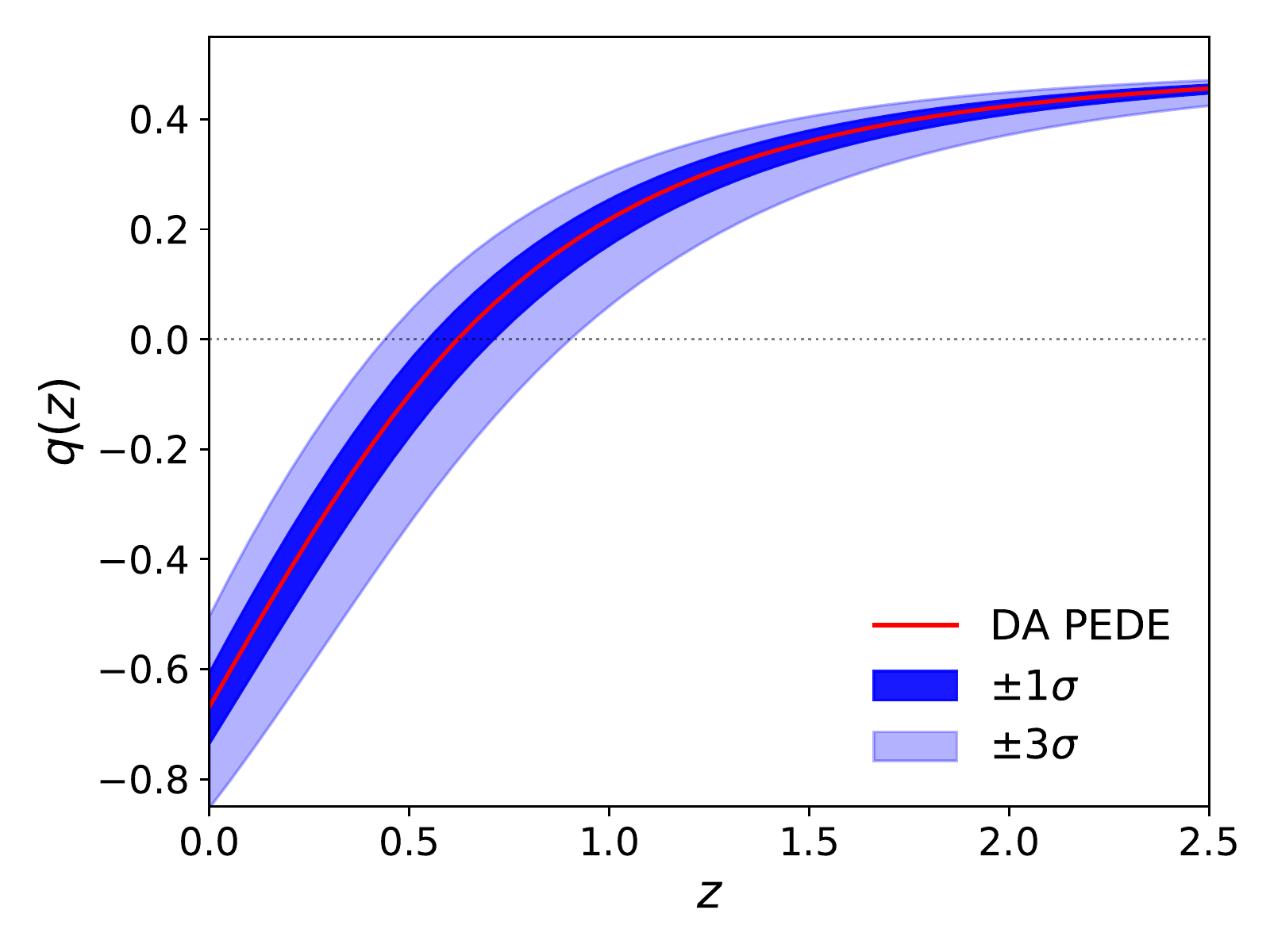} \\
\includegraphics[width=5.5cm,scale=0.55]{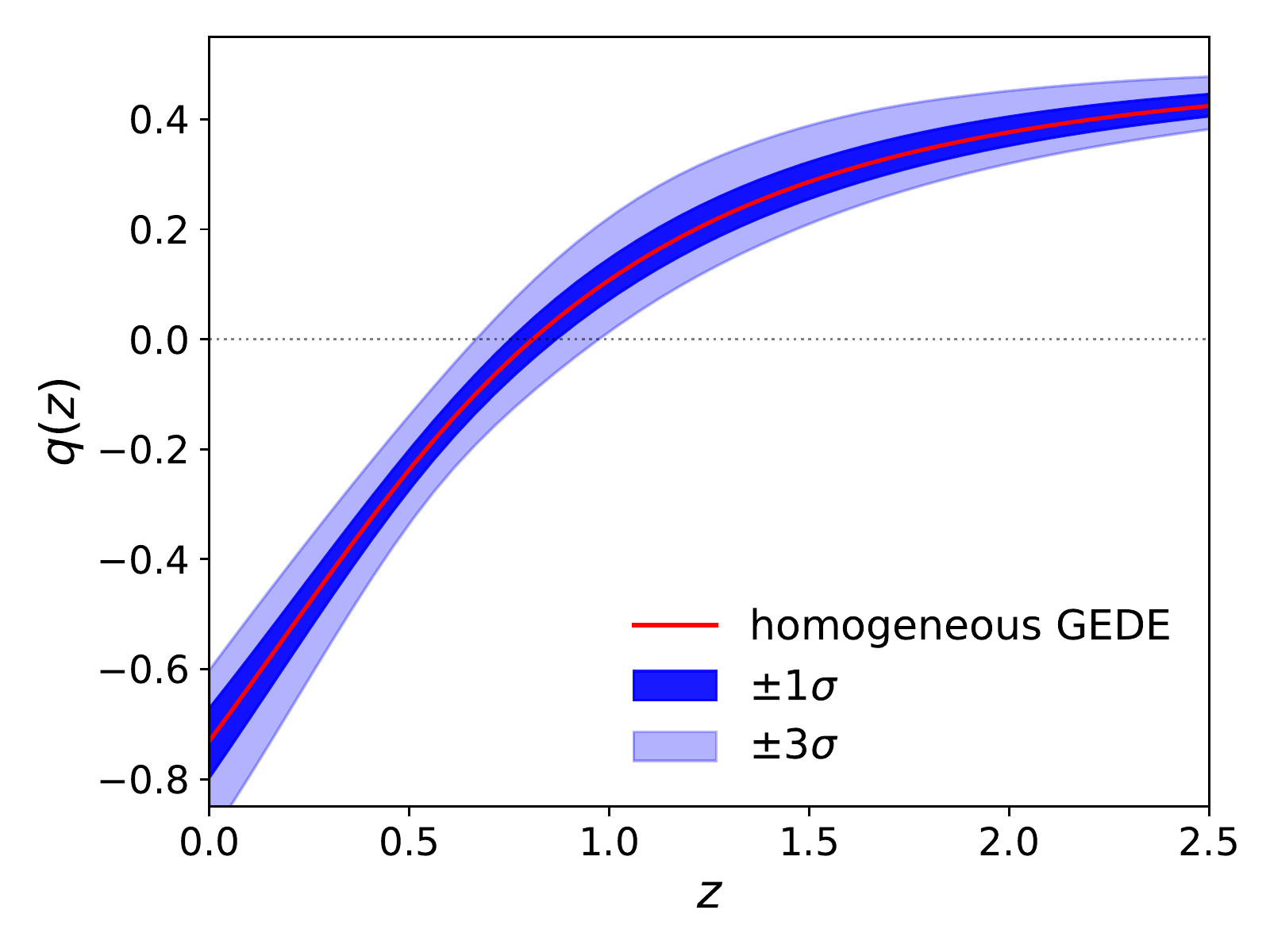} 
\includegraphics[width=5.5cm,scale=0.55]{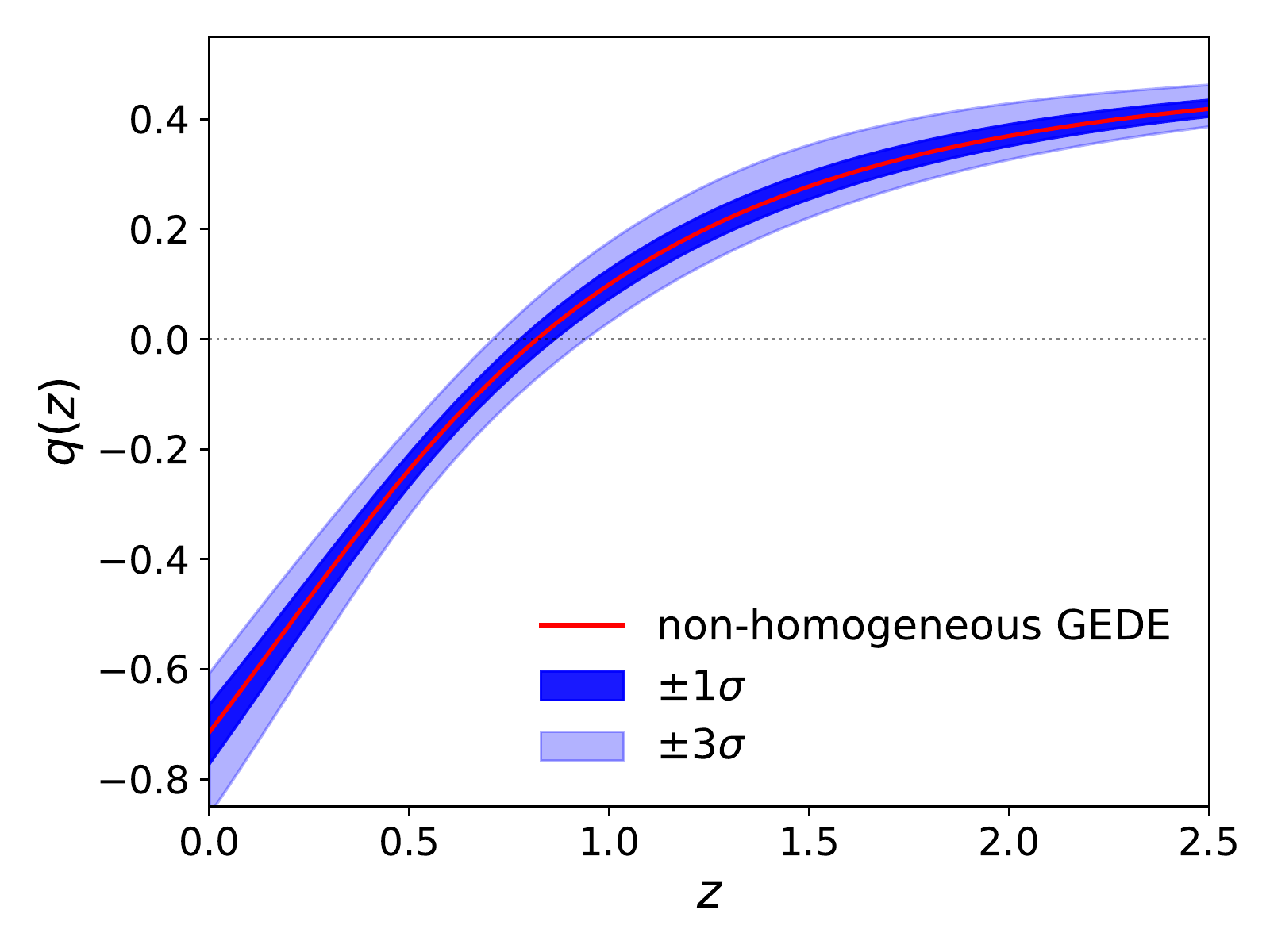} 
\includegraphics[width=5.5cm,scale=0.55]{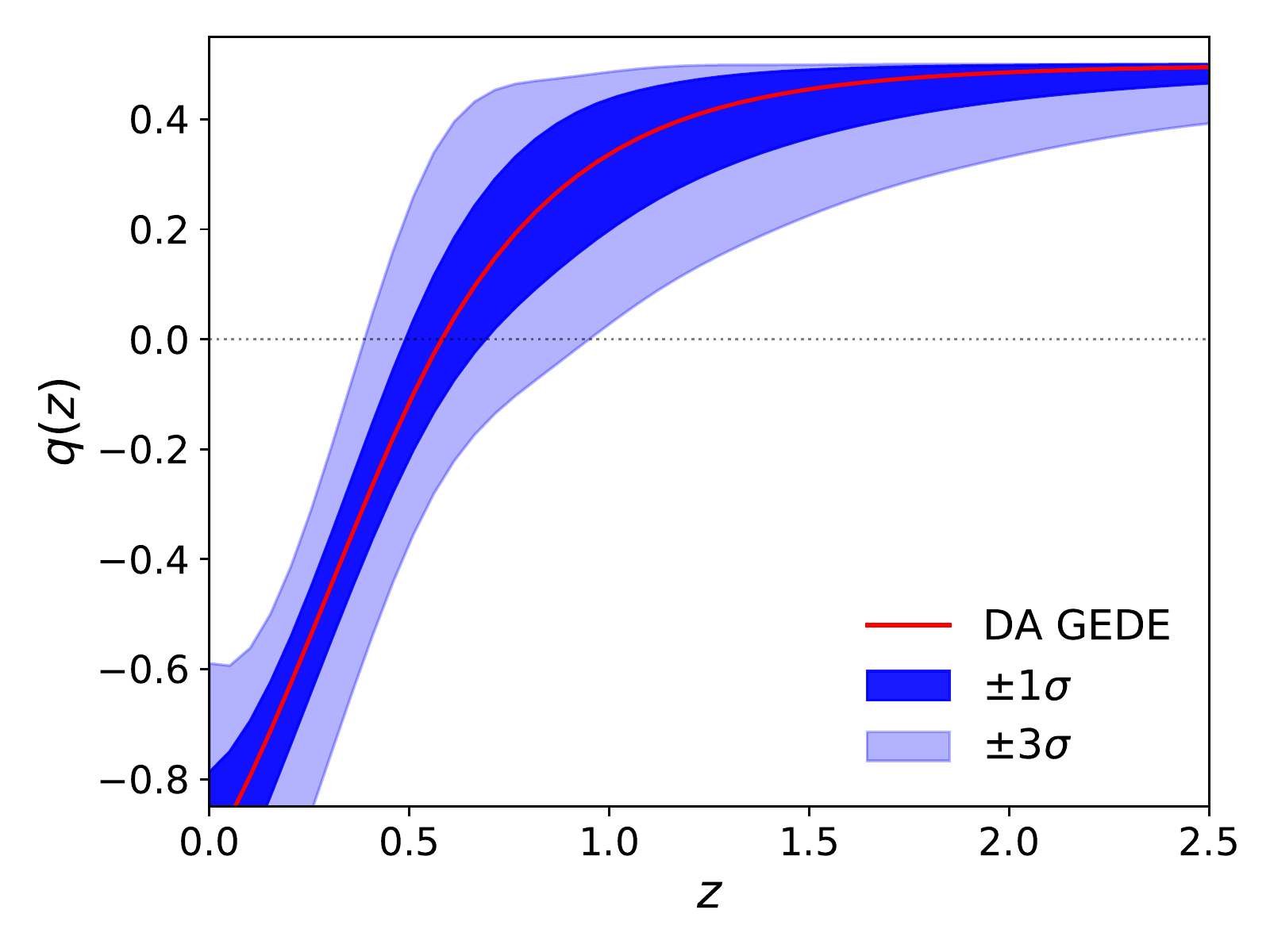} \\
\caption{Reconstruction of the deceleration parameter for PEDE (top panel) and GEDE (bottom panel) using the mean values constraints from the (non-)homogeneous and DA OHD samples at left, middle and right side respectively of the panel. The darker (lighter) band represents the uncertainty at $1\sigma$ ($3\sigma$) CL.}
\label{fig:qz}
\end{figure*}
\begin{figure*}
\centering
\includegraphics[width=5.5cm,scale=0.55]{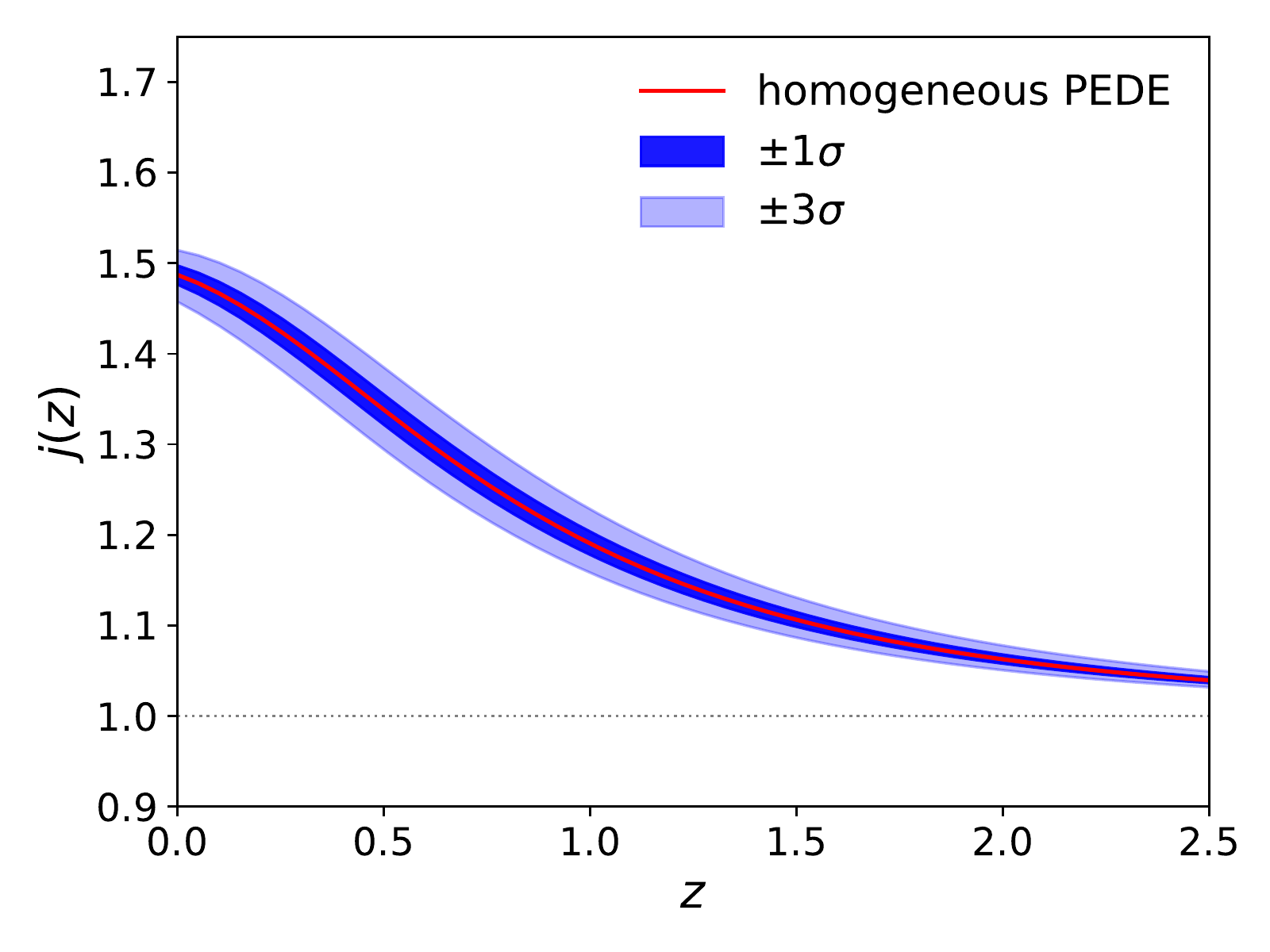} 
\includegraphics[width=5.5cm,scale=0.55]{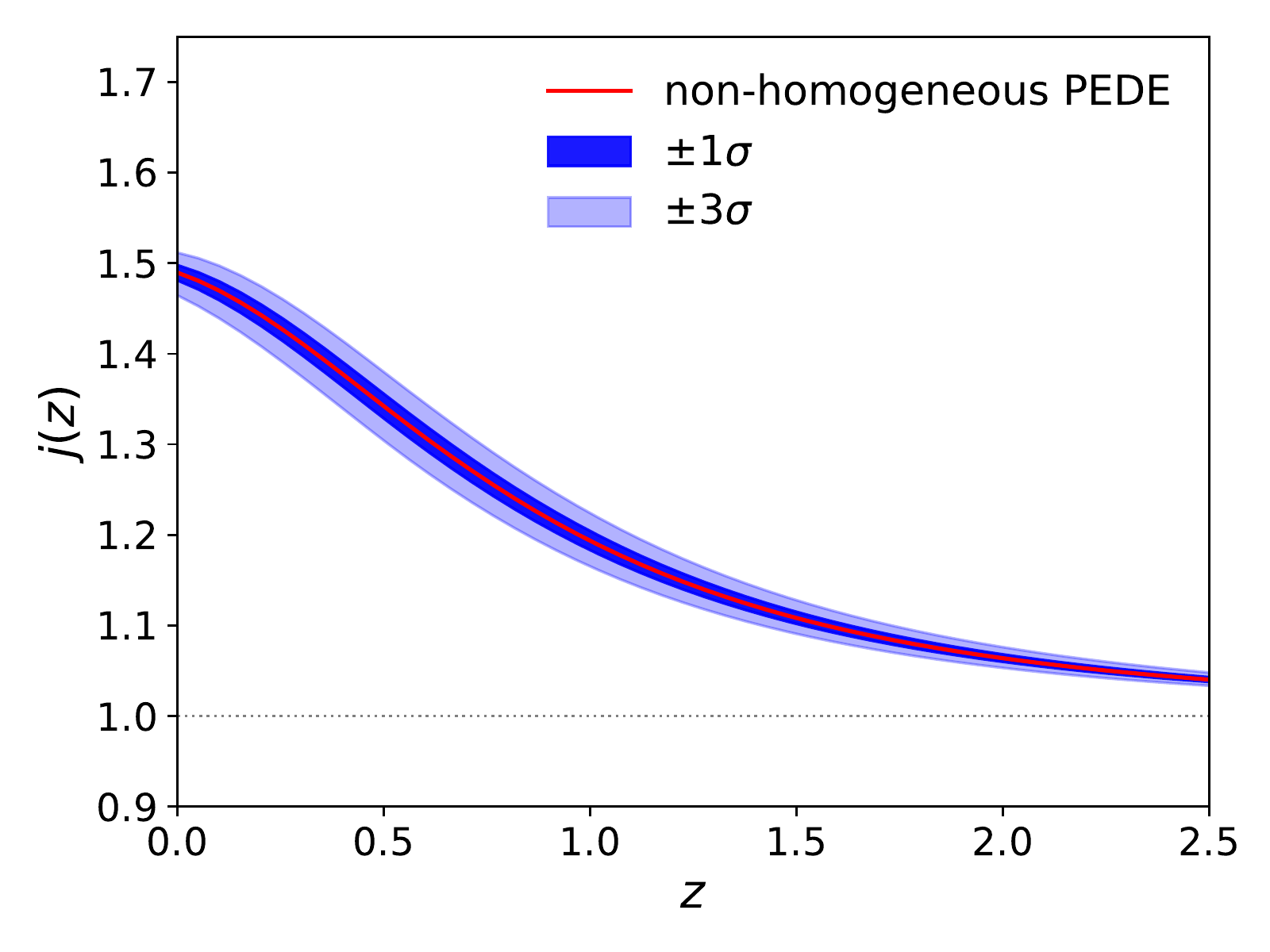} 
\includegraphics[width=5.5cm,scale=0.55]{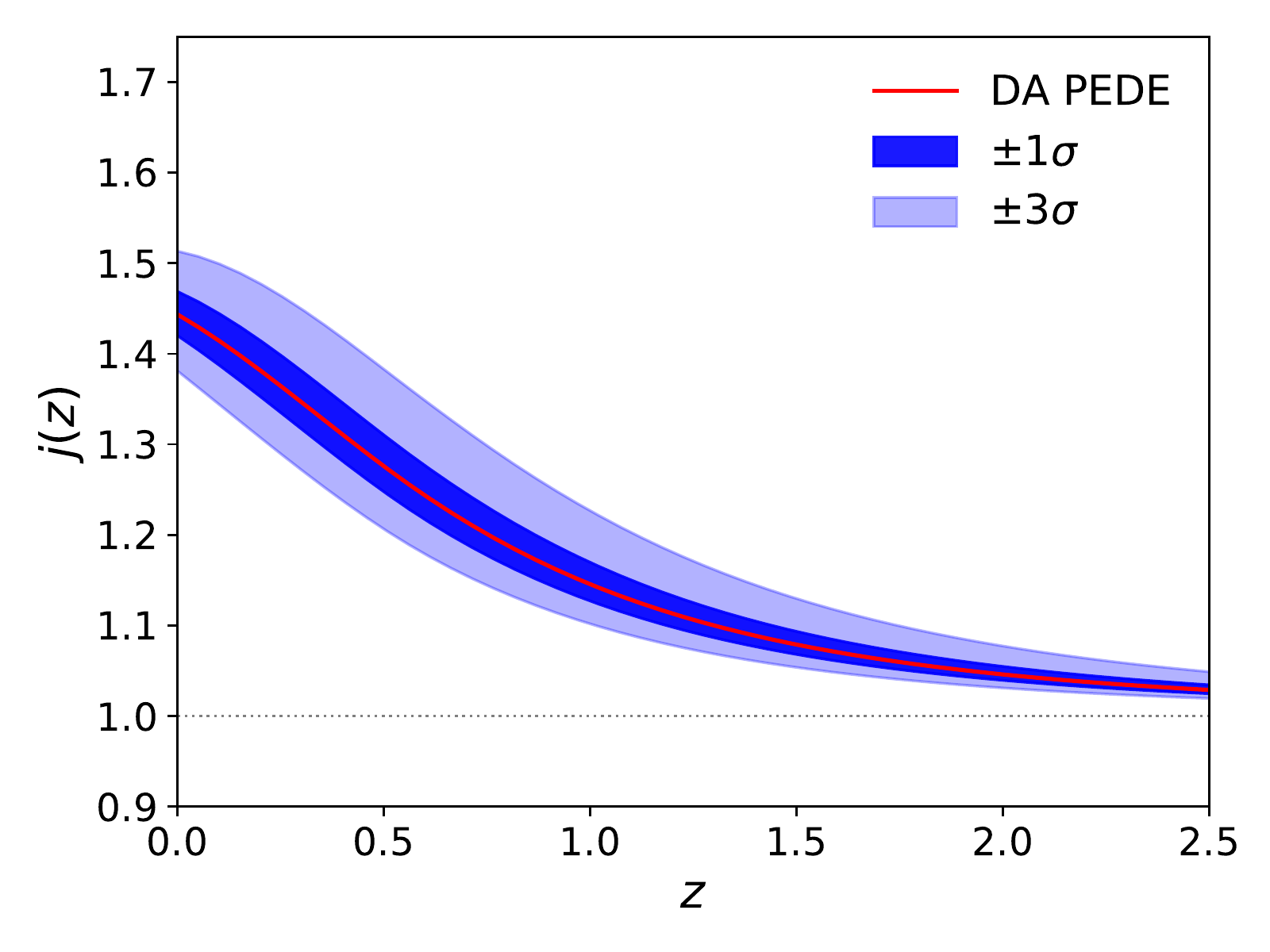} \\
\includegraphics[width=5.5cm,scale=0.55]{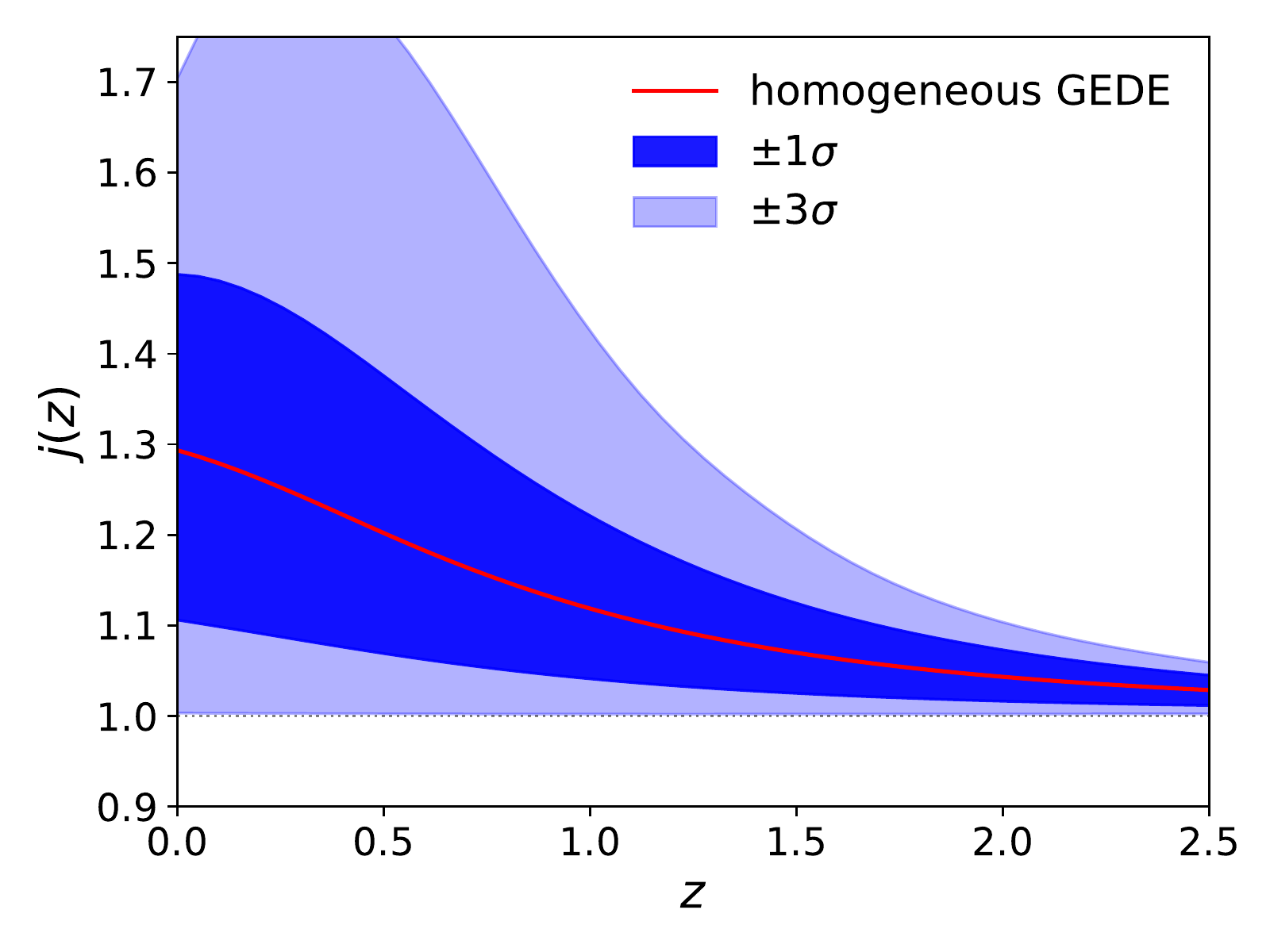} 
\includegraphics[width=5.5cm,scale=0.55]{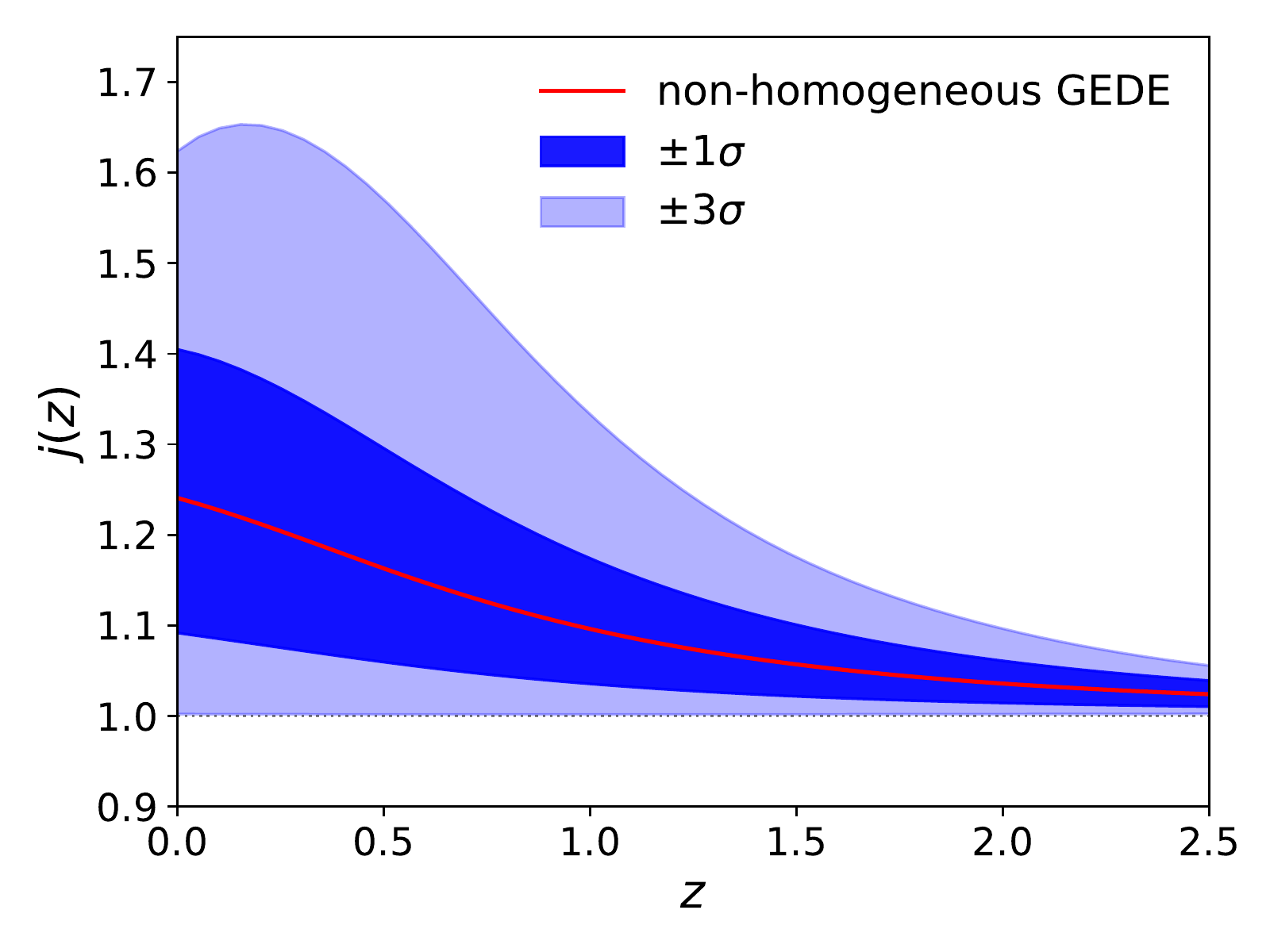} 
\includegraphics[width=5.5cm,scale=0.55]{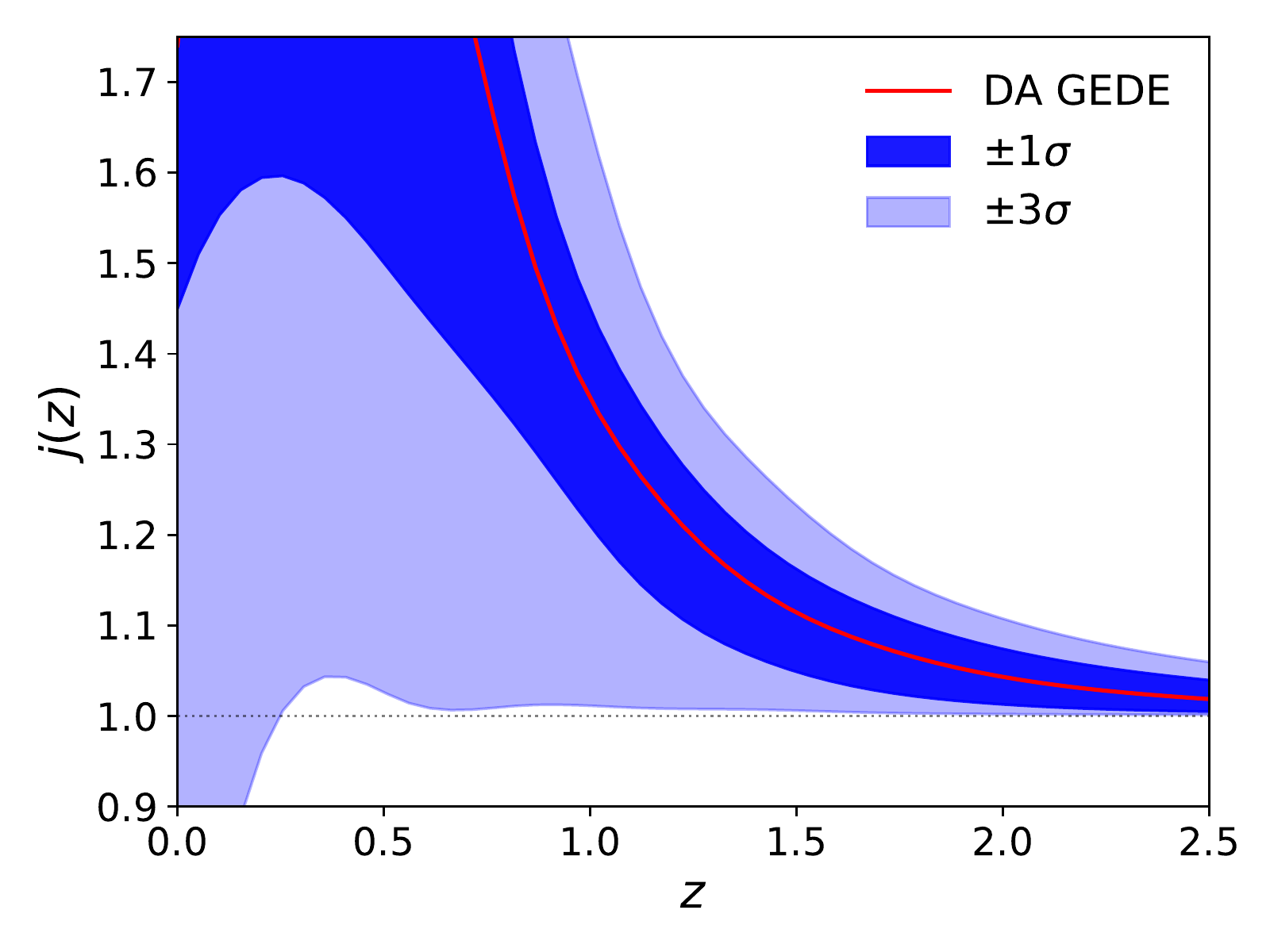} \\
\caption{Reconstruction of the jerk parameter for PEDE (top panel) and GEDE (bottom panel). using the mean values constraints from the (non-)homogeneous and DA OHD samples at left, middle and right side respectively. The darker (lighter) band represents the uncertainty at $1\sigma$ ($3\sigma$) CL.}
\label{fig:jz}
\end{figure*}

Figure \ref{fig:qz} shows the reconstruction of the deceleration parameter as a function of redshift for both, PEDE and GEDE models when the non homogeneous, homogeneous and DA OHD are employed. The universe undergoes a transition from decelerated to accelerated expansion at redshift $0.784^{+0.044}_{-0.044}$ and $0.809^{+0.057}_{-0.057}$ for the PEDE and GEDE models respectively (homogeneous OHD). It is worth to mention that  we observe an earlier deceleration-acceleration transition (close to $0.5$) for DA OHD than for the previously mentioned sample. 
Our constraints are consistent at $1.95\sigma$ and $1.1\sigma$ respectively with the results by \citet{Jesus:2018JCAPJ}. 
Additionally, the reconstruction of the jerk parameter for both models is shown in Figure \ref{fig:jz}. By construction the PEDE and GEDE are DDE models, hence the jerk evolves as a function of the scale factor and it is not equal to one as in the cosmological constant paradigm. 
We  also report the deceleration and jerk parameters at $z=0$ for PEDE as 
$q_0 = -0.784^{+0.028}_{-0.027},\,\, -0.784^{+0.028}_{-0.027}, \;\; -0.668^{+0.061}_{-0.067}$ and 
$j_0 =  1.241^{+0.164}_{-0.149},\,\,  1.487^{+0.010}_{-0.011},\;\; 1.443^{+0.025}_{-0.023}$ using homogeneous, non-homogeneous and DA OHD, respectively. Similarly, for GEDE we estimate 
$q_0 = -0.730^{+0.059}_{-0.067}, \,\,-0.715^{+0.050}_{-0.058},\;\;-0.937^{+0.150}_{-0.151}$ and 
$j_0 =  1.293^{+0.194}_{-0.187}, \,\, 1.241^{+0.164}_{-0.149},\;\; 1.741^{+0.168}_{-0.291}$ when homogeneous, non-homogeneous and DA OHD are considered. We found that the estimate of $q_0$($j_0$), using DA sample, is consistent within $1.9\sigma$($1.5\sigma$)  with the previous values using the (non-) homogeneous samples. 

%%%%%%%%%%%%%%%%%%%%%%%%%%%%%%%%%%%%%%%%%
\section{Dynamical system analysis}\label{sec:stability}
%%%%%%%%%%%%%%%%%%%%%%%%%%%%%%%%%%%%%%%%%

In this section, we investigate the PEDE and GEDE models from the dynamical system approach to obtain the critical points and stability conditions of the models.
This phase-space and stability examination let us to bypass the non-linearities of the cosmological equations, and facilitates a complete analytical
treatment, to obtain a qualitative description of the global dynamics of
these scenarios, which is independent of the initial conditions and the
specific evolution of the universe. Furthermore, in these asymptotic
solutions we are able to calculate various observable quantities, such as the DE
and total equation-of-state parameters, the deceleration parameter,
the density parameters for the different species, etc., that allows us to classify the solution. 

In order to perform the stability analysis of a given cosmological scenario,
one first transforms it to its autonomous form $\label{eomscol}
\textbf{X}'=\textbf{f(X)}$
\citep{Ellis,Ferreira:1997au,Copeland:1997et,Perko,Coley:2003mj,Copeland:2006wr,Chen:2008ft,Cotsakis:2013zha,Giambo:2009byn},
where $\textbf{X}$ is a column vector containing some auxiliary variables and primes denote derivative
with respect to a time variable (conveniently chosen). Then, one extracts the  critical points
$\bf{X_c}$  by imposing the condition  $\bf{X}'=0$, and in order to determine
their stability properties, one expands around them with $\textbf{U}$ the
column vector of the perturbations of the variables. Therefore,
for each critical point the perturbation equations are expanded to first
order as $\label{perturbation} \textbf{U}'={\bf{Q}}\cdot
\textbf{U}$, with the matrix ${\bf {Q}}$ containing the coefficients of the
perturbation equations. The eigenvalues of ${\bf {Q}}$ determine the type and
stability of the specific critical point.

%%%%%%%%%%%%%%%%%%%%%%%%%%%%%%%%%%%%
\subsection{PEDE model}
%%%%%%%%%%%%%%%%%%%%%%%%%%%%%%%%%%%%
To start our analysis, it is convenient to write the 
cosmic evolution equations in terms of the scale factor.
Using the rule  
\begin{equation}
    \frac{d \rho_i}{d t}=\frac{d \rho_i}{d a} \frac{d a}{d t}= a H \frac{d \rho_i}{d a},
\end{equation} and using units where $8 \pi G=1$, 
the field equations are written as
\begin{small}
\begin{subequations}
\label{non_min_2}
	\begin{eqnarray}
	&&\rho_{DE}'(a)+3 (1+ w(a))\frac{\rho_{DE}(a)}{a}=0,\label{Fried1}\\
	&&\rho_{\rm{m}}'(a)+3 \frac{\rho_{\rm{m}}(a)}{a}=0,\label{matter1}\\
	&&\rho_{r}'(a)+4 \frac{\rho_{r}(a)}{a}=0, \label{rad}\\
	&& \frac{{H}'(a)}{H(a)}=-\frac{3}{2}\left(1+w(a)\right)\frac{\Omega_{DE} }{a}-\frac{3}{2}\frac{\Omega_{\rm{m}}}{a}-2\frac{\Omega_{r}}{a},\label{Rachdx}\\
	&&3H^2(a)=\rho_{DE}(a)+\rho_{\rm{m}}(a)+\rho_{r}(a).\label{Fried2}
	\end{eqnarray}
\end{subequations}
\end{small}
Integrating \eqref{Fried1} with the EoS $w(a)$ given by
	\begin{equation}
    w(a)=\,-\frac{1}{3 {\rm{\ln}}\, 10} \left({1-{\rm{tanh}}\left[{\log}_{10}\,a\right]}\right) -1, 
\end{equation} and considering $\rho_{DE}^{(0)}=\rho_{DE}|_{a=1}=3 H_0^2 \Omega_{\rm{DE}}^{(0)}$ we obtain 
\begin{equation}
    \rho_{DE}(a)= 3 H_0^2  \Omega_{\rm{DE}}^{(0)} \left(\tanh \left(\log_{10}(a)\right)+1\right).
\end{equation}

Hence,  
\begin{equation} \label{eq:DV2}
{\Omega}_{\rm{DE}}\,=\, \frac{H_0^2}{H^2} (1-\Omega_{m}^{(0)}-\Omega_{r}^{(0)})\left[ 1 + {\rm{tanh}}\left( {\log}_{10} a \right) \right].
\end{equation}
Defining the time variable 
$\tau={\log}_{10}\,a$,
we have
$\frac{d f}{d \tau}= \ln (10) a \frac{d f}{d a}$.
Alternatively, we can define the time derivative
$\frac{d f}{d\bar{\tau}}=\frac{H_0^2}{(H_0+ H)^2}\frac{df}{d\tau}$. The new time variable $\bar{\tau}$ can be calculated as a function of the redshift through 
\begin{align}
\label{tauvsz}
 &\frac{d \bar{\tau}}{d z}=-\frac{(1+E(z))^2}{(1+z)\ln 10}\nonumber \\
 & =-\frac{1}{(1+z)\ln 10}\left(1+\left[\Omega_{m}^{(0)}\left(1+z\right)^{3}+\Omega_{r}^{(0)}\left(1+z\right)^{4} \right. \right. \nonumber \\ & \left. \left. +\Omega_{\rm{DE}}^{(0)} \left[1 - {\rm{tanh}}\left( {\log}_{10}(1+z) \right) \right]\right]^{1/2}\right)^2.
\end{align}
Defining
\begin{equation} \label{eq:DV1}
    T= \frac{H_0}{H_0+H}, \; \Omega_{m}=\frac{H_{0}^{2} \Omega_{m}^{(0)}}{a^3 H^2}, \;
\Omega_{r}=\frac{H_{0}^{2} \Omega_{r}^{(0)}}{a^{4} H^2},
\end{equation}
$E(z)$ is related to $T(z)$ by 
\begin{eqnarray}
E(z) = \frac{H}{H_0}=\frac{1-T}{T}.
\end{eqnarray}
Therefore, 
\begin{eqnarray}
&{\Omega}_{\rm{DE}}\,= \frac{T^2}{(1-T)^2} (1-\Omega_{m}^{(0)}-\Omega_{r}^{(0)})\left[ 1 + {\rm{tanh}}\left( {\log}_{10} a \right) \right].
\end{eqnarray}
On the other hand, due to the flatness condition \eqref{eq:flatness} we obtain the restriction 
\begin{equation}
  \frac{1-\Omega_{\rm{m}}-\Omega_{\rm{r}}}{(1-\Omega_{\rm{m}}^{(0)}-\Omega_{\rm{r}}^{(0)})}=  \frac{T^2}{(1-T)^2} \left[ 1 + {\rm{tanh}}\left( {\log}_{10} a \right) \right]. \end{equation}
This implies that the equation of state can be expressed as a function of the phase space variables, that is, 
\begin{small}
	\begin{equation}
	\label{wstate}
    w(T,\Omega_{\rm{m}},\Omega_{\rm{r}})=\,-1-\frac{1}{3 {\rm{\ln}}\, 10} \left[2-\frac{(1-\Omega_{\rm{m}}-\Omega_{\rm{r}})(1-T)^2}{(1-\Omega_{\rm{m}}^{(0)}-\Omega_{\rm{r}}^{(0)})T^2}\right].
\end{equation}
\end{small}
The dynamical system for the vector state $(T, \Omega_{\rm{m}}, \Omega_{\rm{r}})^T$ is now given by 
\begin{small}
\begin{subequations}
\label{EQ.18}
\begin{align}
&\frac{d T}{d \bar{\tau}}=\frac{1}{2} (1-T) T^3 (2 (\Omega_{\rm{m}}+\Omega_{\rm{r}}-1)+\ln (10) (3 \Omega_{\rm{m}}+4 \Omega_{\rm{r}})) \nonumber \\
&+\frac{(1-T)^3 T
   (1-\Omega_{\rm{m}}-\Omega_{\rm{r}})^2}{2(1-\Omega_{\rm{m}}^{(0)}-\Omega_{\rm{r}}^{(0)})},\\
&\frac{d \Omega_{\rm{m}}}{d \bar{\tau}}=T^2 \Omega_{\rm{m}} (\ln (10) (3
   \Omega_{\rm{m}}+4 \Omega_{\rm{r}}-3)+2 (\Omega_{\rm{m}}+\Omega_{\rm{r}}-1))\nonumber \\
   &+\frac{(1-T)^2 \Omega_{\rm{m}}
   (1-\Omega_{\rm{m}}-\Omega_{\rm{r}})^2}{(1-\Omega_{\rm{m}}^{(0)}-\Omega_{\rm{r}}^{(0)})},\\
&\frac{d \Omega_{r}}{d \bar{\tau}}=T^2  \Omega_{\rm{r}} (\Omega_{\rm{m}} (2+3 \ln (10))+2
   (\Omega_{\rm{r}}-1) (1+2\ln (10)))\nonumber \\
   & +\frac{(1-T)^2 \Omega_{\rm{r}}
   (1-\Omega_{\rm{m}}-\Omega_{\rm{r}})^2}{(1-\Omega_{\rm{m}}^{(0)}-\Omega_{\rm{r}}^{(0)})},
  \end{align}
\end{subequations}
\end{small}
\noindent defined on the bounded phase space $
\left\{ (T,\Omega_{\rm{m}}, \Omega_{\rm{r}})\in \mathbb{R}^3: 0\leq T\leq 1, \Omega_{\rm{m}}+ \Omega_{\rm{r}}\leq 1,  \Omega_{\rm{m}}\geq 0, \Omega_{\rm{r}}\geq 0 \right\}$. We have three parameters in the model, $\Omega_{\rm{m}}^{(0)}, \Omega_{\rm{r}}^{(0)},  \Omega_{\rm{DE}}^{(0)}=1-\Omega_{\rm{m}}^{(0)} -  \Omega_{\rm{r}}^{(0)}$, which represent the values of  $\Omega_{\rm{m}}, \Omega_{\rm{r}},  \Omega_{\rm{DE}}$ at redshift $z=0$ ($T=0.5$). For the PEDE model, these parameters are constrained in previous section, for the following qualitative and numerical analysis we take the homogeneous constraints,  $\left(\Omega_{\rm{m}}^{(0)},\Omega_{\rm{r}}^{(0)},\Omega_{\rm{DE}}^{(0)}\right)=\left(0.252,7.62\times 10^{-5},0.747 \right)$, which are less unbiased for any fiducial cosmological model (see \S \ref{sec:data}). Notice that
multiplying term by term the system \eqref{EQ.18} by the equation \eqref{tauvsz}, results in a system which can be integrated in terms of redshift.
\begin{table*}
\caption{\label{tab:1} Stability of the equilibrium points of the  system  \eqref{EQ.18}.}
        \centering
    \begin{tabular}{|cccc|}\hline
    Label & $(T, \Omega_{\rm{m}}, \Omega_{r})$ & Eigenvalues & Stability \\\hline
       $P_1$ & $\left(\frac{1}{1+ \sqrt{2\Omega_{\rm{DE}}^{(0)}}}, 0,  0\right)$ & $\left\{-\frac{2}{\left( \sqrt{2\Omega_{\rm{DE}}^{(0)}}+1\right)^2},-\frac{3 \ln (10)}{\left( \sqrt{2\Omega_{\rm{DE}}^{(0)}}+1\right)^2},-\frac{4 \ln (10)}{\left(\sqrt{2\Omega_{\rm{DE}}^{(0)}}+1\right)^2}\right\} $ & sink \\\hline
       $P_2$ & $\left(\frac{1}{1-\sqrt{2\Omega_{\rm{DE}}^{(0)}}},  0,  0\right)$ & $\left\{-\frac{2}{\left( \sqrt{2\Omega_{\rm{DE}}^{(0)}}-1\right)^2},-\frac{3 \ln (10)}{\left( \sqrt{2\Omega_{\rm{DE}}^{(0)}}-1\right)^2},-\frac{4 \ln (10)}{\left( \sqrt{2\Omega_{\rm{DE}}^{(0)}}-1\right)^2}\right\} $ & sink\\\hline
       $P_3$ & $\left(1, 0, 1 \right)$ & $\{2+4 \ln(10),-2 \ln (10),\ln (10)\}$ & saddle \\\hline
       $P_4$ & $\left( 1, 1,  0\right)$ & $\left\{2+3 \ln(10),-\frac{3 \ln (10)}{2},-\ln (10)\right\}$ & saddle \\\hline
       $P_5$ & $\left(1, 0,  0\right)$ & $\{-2 (1+2 \ln(10)),-2-3 \ln (10),1\}$ & saddle \\\hline
       $P_6$ & $\left(0, 0,  0\right)$ & $\left\{\frac{1}{1-\Omega_{\rm{m}}^{(0)}-\Omega_{\rm{r}}^{(0)}}, \frac{1}{1-\Omega_{\rm{m}}^{(0)}-\Omega_{\rm{r}}^{(0)}}, \frac{1}{2 (1-\Omega_{\rm{m}}^{(0)}-\Omega_{\rm{r}}^{(0)})}\right\}$ & source \\\hline
       $P_7$ & $\left(0,\Omega_{\rm{m}},  1-\Omega_{\rm{m}}\right)$ & $\{0,0,0\}$ & nonhyperbolic\\\hline
       $P_8$ & $\left(0,  0,  1\right)$ & $\{0,0,0\} $ & nonhyperbolic\\\hline
       $P_9$ & $\left(0, 1,  0\right)$ & $\{0,0,0\}$ & nonhyperbolic\\\hline\end{tabular}
    \end{table*}

We can study the dynamical system \eqref{EQ.18} as we discussed in Table \ref{tab:1}.
\begin{figure}
    \centering
    \includegraphics[width=6cm,scale=0.45]{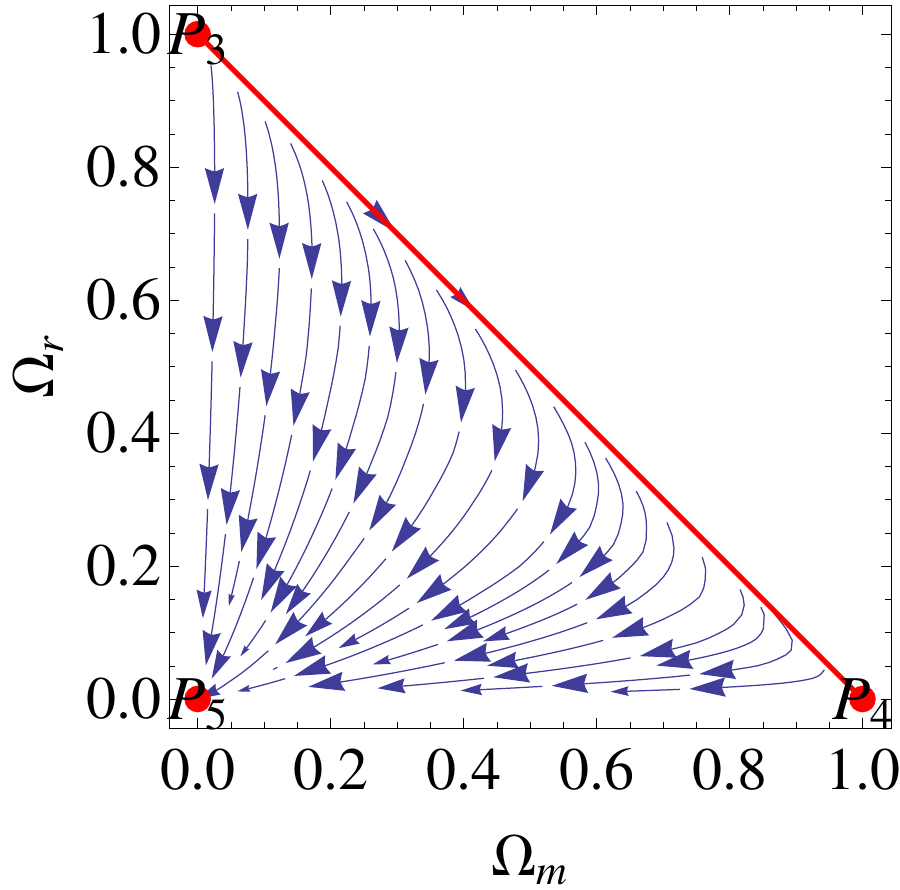}
    \caption{Dynamics of the system \eqref{EQ.18} on the invariant set $T=1$. The equilibrium point $P_3: \left(1, 0,  1\right)$ is a local source,  $P_4: \left(1, 1,  0\right)$ is a saddle and $P_5: \left(1, 0,  0\right)$ is a local sink (but a saddle in the 3D phase space). }
    \label{fig:PEDEphase-plot_T_1}
\end{figure}
\begin{figure}
    \centering
    \includegraphics[width=6cm,scale=0.45]{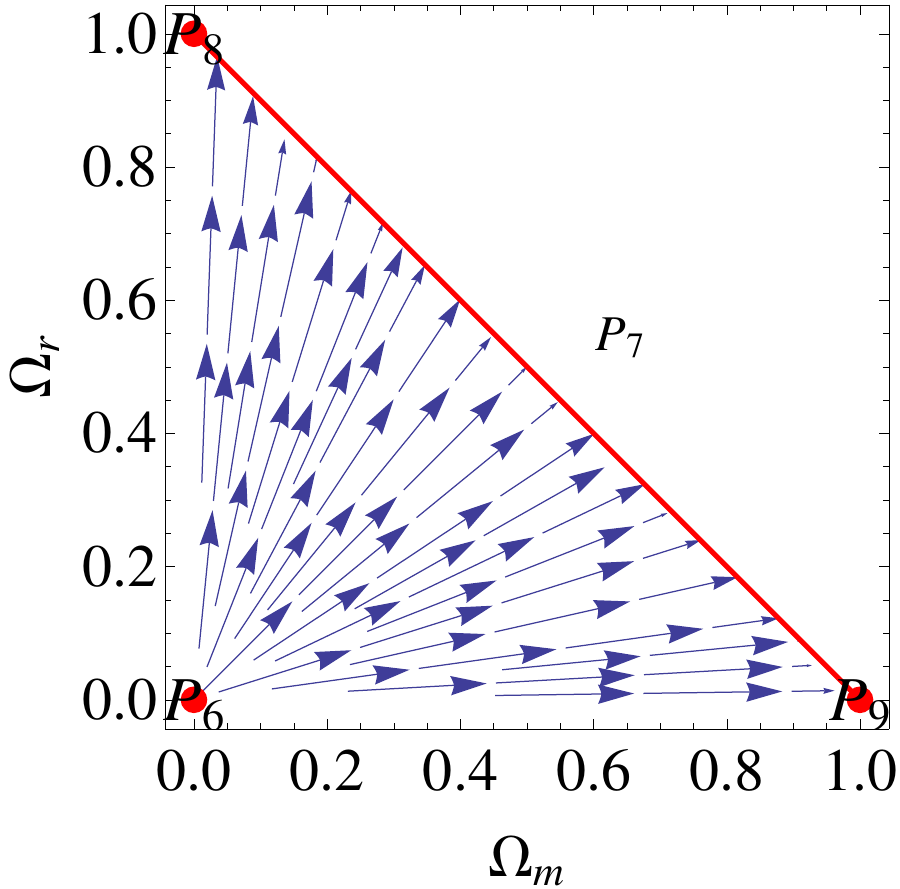}
    \caption{Dynamics of the system \eqref{EQ.18} on the invariant set $T=0$. The line $P_7: \left(0,\Omega_{\rm{m}},  1-\Omega_{\rm{m}}\right)$, and its endpoints $P_8$ and $P_9$ are local attractors. $P_6$ is the global source.}
    \label{fig:PEDEphase-plot_T_0}
\end{figure}
The system \eqref{EQ.18} admits two relevant invariant sets $T=1$ and $T=0$. The variable $T$ satisfies $T\rightarrow 0$ when $H\rightarrow \infty$;  $T\rightarrow 1$ when $H\rightarrow 0$; and $T=0.5$ when $H\rightarrow H_0$.
In the invariant set $T=1$ the dynamics of the system \eqref{EQ.18} is as shown in Fig. \ref{fig:PEDEphase-plot_T_1}. The equilibrium point $P_3: \left(1, 0,  1\right)$ is a local source,  $P_4: \left(1, 1,  0\right)$ is a saddle and $P_5: \left(1, 0,  0\right)$ is a local sink (but a saddle in the 3D phase space).
On the other hand, the dynamics at the invariant set $T=0$ is governed by an integrable 2D dynamical system 
such that the orbit passing through $(T, \Omega_{\rm{m}}, \Omega_{\rm{r}})=(0, \Omega_{\rm{m},0}, \Omega_{\rm{r},0})$ at $\bar{\tau}=\bar{\tau}_0$ is given by 
\begin{equation}
    \Omega_{\rm{r}}(\Omega_{\rm{m}})= \frac{\Omega_{\rm{m}} \Omega_{\rm{r},0}}{\Omega_{\rm{m},0}}. 
\end{equation}
For this solution, the relation between $\bar{\tau}$ and $\Omega_{\rm{m}}$ is
\begin{align}
   &\bar{\tau} (\Omega_{\rm{m}})= \bar{\tau}_0+\frac{(\Omega_{\rm{m}}-\Omega_{\rm{m},0}) (1-\Omega_{\rm{m}}^{(0)}-\Omega_{\rm{r}}^{(0)}) (1-\Omega_{\rm{m},0}-\Omega_{\rm{r},0})}{(\Omega_{\rm{m},0}-\Omega_{\rm{r},0})
   ((1-\Omega_{\rm{m}}) \Omega_{\rm{m},0}-\Omega_{\rm{m}} \Omega_{\rm{r},0})} \nonumber \\
   & +(1-\Omega_{\rm{m}}^{(0)}-\Omega_{\rm{r}}^{(0)}) \ln \left(\frac{\Omega_{\rm{m}} (1-\Omega_{\rm{m},0}-\Omega_{\rm{r},0})}{(1-\Omega_{\rm{m}}) \Omega_{\rm{m},0}-\Omega_{\rm{m}} \Omega_{\rm{r},0}}\right).
\end{align}
Figure \ref{fig:PEDEphase-plot_T_0} illustrates the dynamics of the system \eqref{EQ.18} on the invariant set $T=0$. The line $P_7: \left(0,\Omega_{\rm{m}},  1-\Omega_{\rm{m}}\right)$, and the endpoints $P_8$ and $P_9$ are local attractors.  $P_6$ is the source 
($\bar{\tau}$ was re scaled by the factor $1-\Omega_{\rm{m}}^{(0)}-\Omega_{\rm{r}}^{(0)}>0$). \\
In the 3D phase space, the late-time attractors are the equilibrium points $P_{1,2}$ with $T= \frac{1}{1\pm \sqrt{2 \Omega_{\rm{DE}}^{(0)}}}, \Omega_{\rm{m}}=0, \Omega_{\rm{r}}=0$. Therefore $H_{\pm}=\pm  \sqrt{2 \Omega_{\rm{DE}}^{(0)}} H_0$. The corresponding cosmological solutions are $a_{\pm}(t)= a_0 e^{\pm  \sqrt{2 \Omega_{\rm{DE}}^{(0)}} H_0 t}$. The choice $+$, that corresponds to $P_1$, belongs to an ever expanding de Sitter solution. The solution corresponding to $P_2$ satisfies $a\rightarrow 0$ at late times; an static solution. However, this solution is not physical because the condition $T\geq 0$ requires $0\leq \Omega_{\rm{DE}}^{(0)}<\frac{1}{2}$, which is not supported (at $> 5\sigma$) neither by the narrow bound placed by Planck data $\Omega_{DE}^{(0)}=0.6889\pm0.0056$ \citep{Planck2018}, nor by our values $\Omega_{DE}^{(0)}=0.748^{+0.016}_{-0.015}$ (homogenous OHD), $\Omega_{\rm{DE}}^{(0)}=0.6801^{+0.039}_{-0.036}$ (DA OHD).
\\
There are three solutions $P_3$, $P_4$ and $P_5$ dominated by radiation, DM and DE, respectively, that satisfy $T=1$. This means that $H=0$ for these solutions, and they are saddles.  
\\
The point $P_6$ is the source, it satisfies $\Omega_{\rm{m}}=0, \Omega_{\rm{r}}=0$, therefore, it is dominated by DE. As $T=0$, this implies that $H\rightarrow \infty$. Because it is a source, it represents the initial stages of the cosmic evolution, dominated by DE. This means that for the model not only dark energy accounts for the recent accelerated phase of the evolution  but also the initial stage is driven by an accelerated dark-energy dominated expanding phase.  
\\
To analyse the nonhyperbolic points $P_7$, $P_8$ and $P_9$ that satisfy $T\rightarrow 0$, we rely on numerical examination, where we see that they behave as saddles as shown in the top of Fig. \ref{fig:3D}.
However, when the dynamics is restricted to the invariant set  $T=0$, it is governed by an integrable 2D dynamical system, such that the line $P_7: \left(0,\Omega_{\rm{m}},  1-\Omega_{\rm{m}}\right)$, along with the endpoints $P_8$ and $P_9$, are local attractors (as shown in Fig. \ref{fig:PEDEphase-plot_T_0}), whereas $P_6$ is the global source.

%%%%%%%%%%%%%%%%%%%%%%%%%%%%%%%%%%%%%%
\subsection{GEDE model}
%%%%%%%%%%%%%%%%%%%%%%%%%%%%%%%%%%%%%%

In this section we investigate the  GEDE model with $\Omega_{DE}$ given by Eq. \eqref{eq:omegade_gene}
whose evolution is given by \eqref{non_min} with $w(z)$ defined by \eqref{wPEDE}.
\\
Due to the flatness condition given by Eq. \eqref{eq:flatness} we obtain the restriction 
\begin{equation}
  \frac{1-\Omega_{\rm{m}}-\Omega_{\rm{r}}}{(1-\Omega_{\rm{m}}^{(0)}-\Omega_{\rm{r}}^{(0)})}=  \frac{T^2}{(1-T)^2} \left[\frac{ 1 - {\rm{tanh}}\left( {\Delta \log}_{10}(\frac{1+z}{1+z_{t}}) \right)}{ 1 + {\rm{tanh}}\left(\Delta {\log}_{10}(1+z_{t}) \right)}\right]. \end{equation}
  This implies that the equation of state can be expressed as a function of the phase space variables, that is, 
 \begin{small}
    \begin{align}
    \label{EQ.31}
&w(T,\Omega_{\rm{m}},\Omega_{\rm{r}})  =\nonumber \\
 & \,-1-\frac{\Delta  \left(2-\frac{(T-1)^2 (\text{$\Omega
   $m}+\Omega_{\rm{r}}-1) \left(\tanh \left(\frac{\Delta 
   \ln (z_t+1)}{\ln (10)}\right)+1\right)}{T^2
   (\Omega_{\rm{m}}^{(0)}+\Omega_{\rm{r}}^{(0)}-1)}\right)}{3 \ln
   (10)}. 
\end{align}
\end{small}
In this case  we calculate $\bar{\tau}$ as a function of the redshift through 
\begin{align}
\label{tauvsz2}
 &\frac{d \bar{\tau}}{d z}=-\frac{(1+E(z))^2 }{(1+z)\ln 10}\nonumber \\
 & =-\frac{1}{(1+z)\ln 10}\left(1+\left[\Omega_{m}^{(0)}\left(1+z\right)^{3}+\Omega_{r}^{(0)}\left(1+z\right)^{4} \right. \right. \nonumber \\
& \left. \left.  +\Omega_{\rm{DE}}^{(0)} \frac{ 1 - {\rm{tanh}}\left( {\Delta \log}_{10}(\frac{1+z}{1+z_{t}}) \right)}{ 1 + {\rm{tanh}}\left(\Delta {\log}_{10}(1+z_{t}) \right)} \right]^{1/2}\right)^2.
\end{align}
\\
The dynamical system for the vector state $(T, \Omega_{\rm{m}}, \Omega_{r})^T$ is now given by
\begin{small}
\begin{subequations}
\label{EQ.32}
\begin{align}
&\frac{d T}{d \bar{\tau}}=-\frac{1}{2} (T-1) T^3 (2
   \Delta  (\Omega_{\rm{m}}+\Omega_{\rm{r}}-1)+\ln (10) (3 \Omega_{\rm{m}}+4 \Omega_{\rm{r}})) \nonumber \\
   & +\frac{\Delta  (T-1)^3 T (\Omega_{\rm{m}}+\Omega_{\rm{r}}-1)^2 g(\Delta, z_t)}{2 (\Omega_{\rm{m}}^{(0)}+\Omega_{\rm{r}}^{(0)}-1)},\\
&\frac{d \Omega_{\rm{m}}}{d \bar{\tau}}=T^2 \Omega_{\rm{m}} (2 \Delta  (\Omega_{\rm{m}}+\Omega_{\rm{r}}-1)+\ln (10) (3 \Omega_{\rm{m}}+4 \Omega_{\rm{r}}-3))\nonumber \\
   & +\frac{\Delta  (1-T)^2 \Omega_{\rm{m}} (1-\Omega_{\rm{m}}-\Omega_{\rm{r}})^2 g(\Delta, z_t))}{1-\Omega_{\rm{m}}^{(0)}-\Omega_{\rm{r}}^{(0)}},\\
&\frac{d \Omega_{\rm{r}}}{d \bar{\tau}}=T^2 \Omega_{\rm{r}}(2 \Delta  (\Omega_{\rm{m}}+\Omega_{\rm{r}}-1)+\ln (10)(3\Omega_{\rm{m}}+4 \Omega_{\rm{r}}-4))\nonumber \\
   & +\frac{\Delta  (1-T)^2 \Omega_{\rm{r}} (1-\Omega_{\rm{m}}-\Omega_{\rm{r}})^2 g(\Delta, z_t)}{1-\Omega_{\rm{m}}^{(0)}-\Omega_{\rm{r}}^{(0)}},
  \end{align}
\end{subequations}
\end{small}
with 
\begin{equation}
    g(\Delta, z_t)=\tanh \left(\Delta  \log_{10} (z_t+1)\right)+1,
\end{equation} 
defined on the bounded phase space $\left\{ (T,\Omega_{\rm{m}}, \Omega_{\rm{r}})\in \mathbb{R}^3: 0\leq T\leq 1, \Omega_{\rm{m}}+ \Omega_{\rm{r}}\leq 1,  \Omega_{\rm{m}}\geq 0, \Omega_{\rm{r}}\geq 0 \right\}$. In the GEDE model, we take as the observable parameters the homogeneous constraints (which are less unbiased due to any underlying cosmology, see \S \ref{sec:data}): $\left(\Omega_{\rm{m}}^{(0)},\Omega_{\rm{r}}^{(0)},\Omega_{\rm{DE}}^{(0)}\right)=(0.247, 7.72\times 10^{-5},0.752)$.
\begin{table*}
   \caption{\label{tab:3} Stability of the equilibrium points of the system \eqref{EQ.32}. We use the notations  $g(\Delta, z_t)=\tanh \left(\frac{\Delta  \ln (z_t+1)}{\ln (10)}\right)+1,$ and $\tilde{\Lambda}\equiv\sqrt{\frac{2 \Omega_{\rm{DE}}^{(0)}}{ g(\Delta, z_t)}}$.}
        \centering
\begin{tabular}{|cccc|}\hline
 Label & $(T, \Omega_{\rm{m}}, \Omega_{r})$ & Eigenvalues & Stability \\\hline
  $P_1$ & $\left(\frac{1}{1+\tilde{\Lambda}}, 
  0, 0\right)$ & $\left\{-\frac{4 \ln (10)}{(\tilde{\Lambda} +1)^2},-\frac{3 \ln (10)}{(\tilde{\Lambda} +1)^2},-\frac{2 \Delta }{(\tilde{\Lambda} +1)^2}\right\}$ & sink \\\hline
   $P_2$ & $\left(\frac{1}{1-\tilde{\Lambda}}, 0,\  0\right)$ & $\left\{-\frac{4 \ln (10)}{(\tilde{\Lambda} -1)^2},-\frac{3 \ln (10)}{(\tilde{\Lambda} -1)^2},-\frac{2 \Delta }{(\tilde{\Lambda} -1)^2}\right\}$ & sink \\\hline
    $P_3$ & $(1, 0,  1)$ & $\{-2 \ln (10),\ln (10),2 (\Delta +2 \ln (10))\} $ & saddle \\\hline
    $P_4$ & $(1, 1, 0)$ & $\left\{-\frac{3 \ln (10)}{2},-\ln (10),2 \Delta +3 \ln
   (10)\right\}$ & saddle\\\hline
  $P_5$ & $(1,  0, 0)$ & $\{\Delta ,-2 (\Delta +2 \ln (10)),-2 \Delta -3 \ln (10)\}$ & saddle\\\hline    
  $P_6$ & $(0,  0, 0)$ & $\left\{\frac{2 \Delta }{\tilde{\Lambda}^2},\frac{2 \Delta }{\tilde{\Lambda}^2},\frac{\Delta }{\tilde{\Lambda}^2}\right\}$ & source \\\hline

           $P_7$ & $(0,\Omega_{\rm{m}}, 1-\Omega_{\rm{m}})$ & $\{0,0,0\}$ & nonhyperbolic \\\hline
        $P_8$ & $(0, 0, 1)$& $\{0,0,0\} $ & nonhyperbolic \\\hline
        $P_9$ & $(0, 1, 0)$ & $\{0,0,0\}$  & nonhyperbolic \\\hline
     \end{tabular}
    \end{table*}

The stability of the equilibrium points of system \eqref{EQ.32} are discussed in table \ref{tab:3} \footnote{Multiplying term by term system \eqref{EQ.32} by equation \eqref{tauvsz2} 
we obtain a system that can be integrated in terms of redshift.}.
\\
The system \eqref{EQ.32} admits the relevant invariant sets $T=1$ and $T=0$. 

In a similar way as for the PEDE model, the upper bounds within $1\sigma$ confidence levels of the parameter mean values are
\begin{small}
    \begin{eqnarray}
    & z_t \sim \left\{\begin{array}{lcl}
0.403+0.058=0.461, &  \text{homogeneous OHD} \\
0.183+0.094=0.277, & \text{DA OHD}  
\end{array}\right. \label{ZT}\\
 & \Delta \sim \left\{\begin{array}{lcl}
0.690+0.624=1.314, &  \text{homogeneous OHD} \\
3.930+2.304=6.234, & \text{DA OHD}  
\end{array}\right. \label{D}
\end{eqnarray}
\end{small}
\\
and the dynamics is qualitatively the same as for the system \eqref{EQ.18}. That is, in the invariant set $T=1$ the equilibrium point $P_3: \left(1, 0,  1\right)$ is a local source,  $P_4: \left(1, 1,  0\right)$ is a saddle and $P_5: \left(1, 0,  0\right)$ is a local sink (but a saddle in the 3D phase space).
On the other hand, the dynamics at the invariant set $T=0$ is governed by an integrable 2D dynamical system, such that the line $P_7: \left(0,\Omega_{\rm{m}},  1-\Omega_{\rm{m}}\right)$, along with the endpoints $P_8$ and $P_9$ are local attractors, whereas $P_6$ is the global source. 
\\
The late-time attractors on the 3D phase space are the equilibrium points $P_{1,2}$ with $T= \frac{1}{\left(1\pm \tilde{\Lambda}\right)}, \Omega_{\rm{m}}=0, \Omega_{\rm{r}}=0$, with $\tilde{\Lambda}\equiv\sqrt{\frac{2 \Omega_{\rm{DE}}^{(0)}}{ g(\Delta, z_t)}}$. Therefore $H_{\pm}=\pm \tilde{\Lambda}$. The cosmological solutions corresponds to $a_{\pm}(t)= a_0 e^{\pm  \tilde{\Lambda} t}$. The choice $+$, that is associated to $P_1$, corresponds to an ever expanding de Sitter solution. The solution corresponding to $P_2$ satisfies $a\rightarrow 0$ at late times. Therefore, it is an static solution.  However, this solution is not physical because the condition $T\geq 0$, requires $0\leq \Omega_{\rm{DE}}^{(0)}<\frac{g(\Delta, z_t)}{2}$, with 
\begin{equation}
g(\Delta, z_t)\sim \left\{\begin{array}{lcl}
1.213046, &  \text{homogeneous OHD} \\
1.90944,& \text{DA OHD}  
\end{array}\right.
\end{equation}
where we have used the upper bounds $z_t$ and $\Delta $ within $1\sigma$ confidence levels given by \eqref{ZT}, \eqref{D}, respectively.
For homogeneous OHD, we conclude that the interval for $\Omega_{DE}^{(0)}$ is not supported by observations, i.e. the narrow bound from Planck data $\Omega_{DE}^{(0)}=0.6889\pm0.0056$ by \cite{Planck2018}. With our value $\Omega_{DE}^{(0)}=0.753^{+0.018}_{-0.017}$, the restriction has less probability to be satisfied. However, for DA OHD this interval becomes $0\leq \Omega_{\rm{DE}}^{(0)}\lesssim 0.954722$. For this set $\Omega_{DE}^{(0)} \approx 1-\Omega_{\rm{m}}^{(0)}= 0.6801^{+0.039}_{-0.036}$, and the point $P_2$ is allowed by the observations. A conservative upper bound $0.954722$ was calculated with the largest values $z_t, \Delta$, but it takes a lower value ($\sim 0.789332$) for the best-fit values.

\begin{figure}
    \centering
    \includegraphics[width=8cm,scale=0.9]{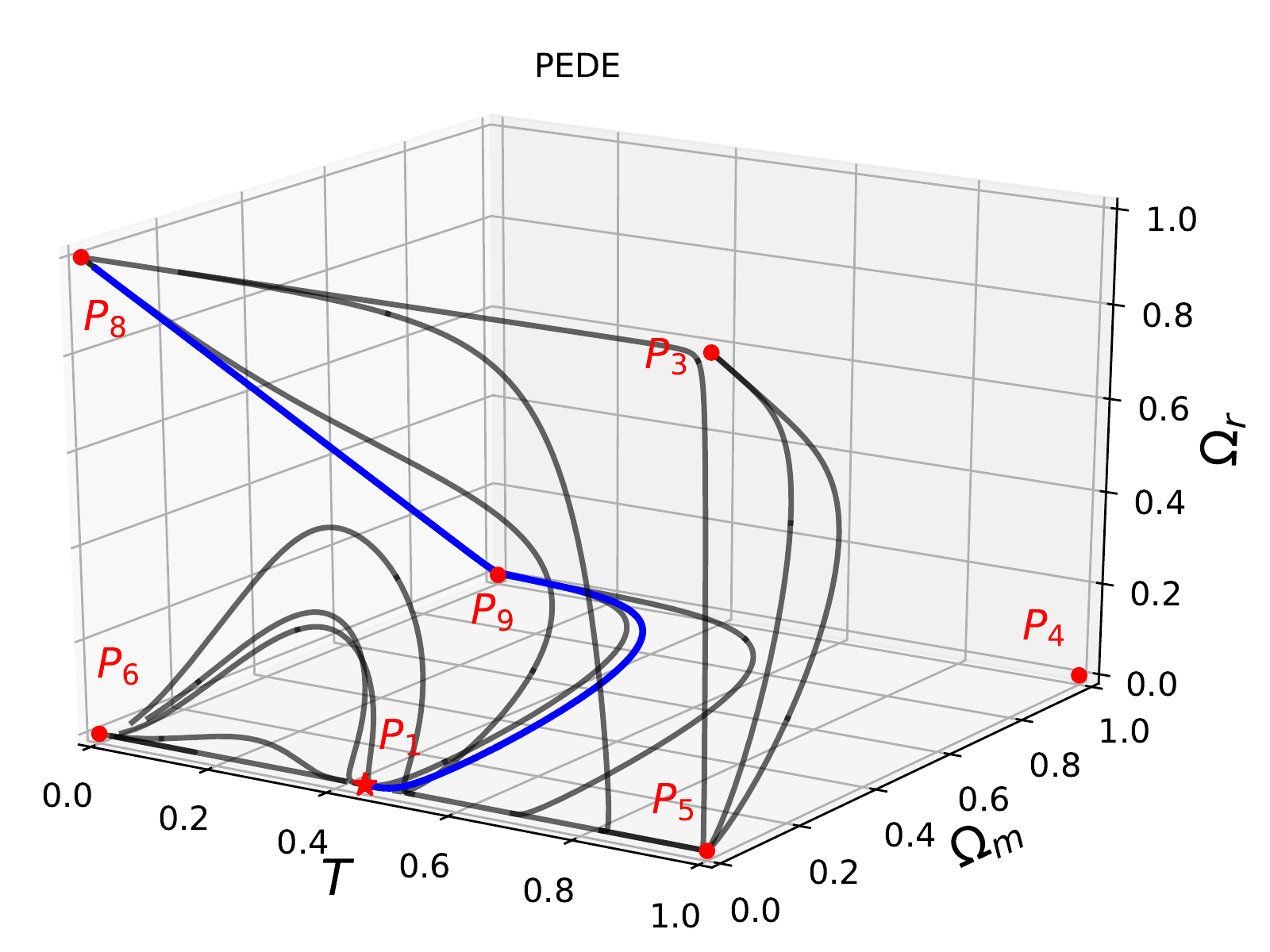}\\
    \includegraphics[width=8cm,scale=0.9]{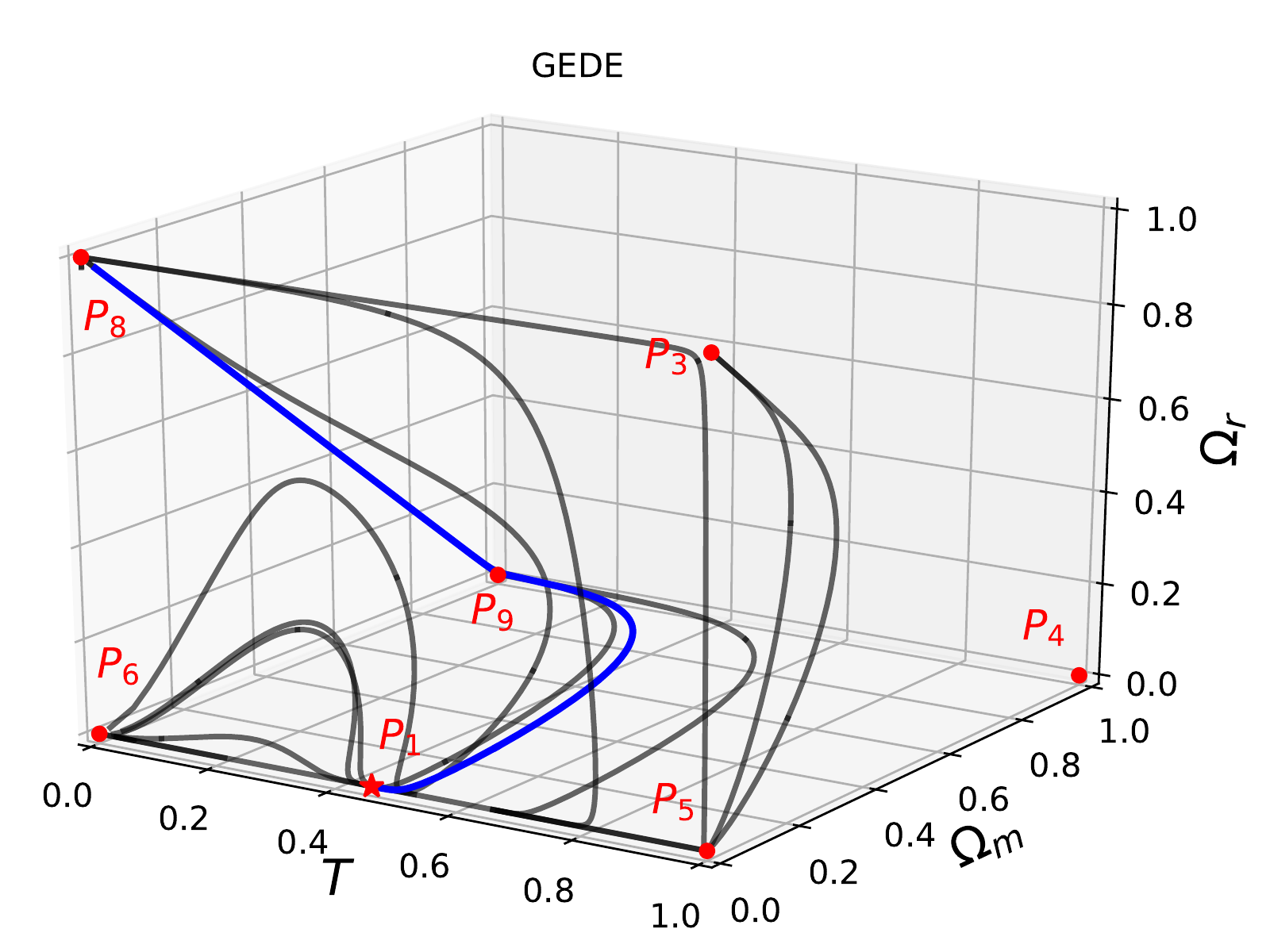}\\
\caption{Dynamics of the systems \eqref{EQ.18} for the PEDE model with $\left(\Omega_{\rm{m}}^{(0)},\Omega_{\rm{r}}^{(0)}\right)=\left(0.252,7.62\times 10^{-5}\right)$ (top panel) and \eqref{EQ.32} for the GEDE model with $\left(\Omega_{\rm{m}}^{(0)},\Omega_{\rm{r}}^{(0)}\right)=\left(0.247, 7.72\times 10^{-5}\right)$ (bottom panel). The blue lines correspond to the orbit with initial condition $(T(0),\Omega_{\rm{m}}{(0)},\Omega_{\rm{r}}{(0)})=(0.5,0.252,7.62\times 10^{-5})$ and $(0.5, 0.247, 7.72\times 10^{-5})$, for PEDE and GEDE respectively, which represents the current universe.   We see that all orbits are attracted by the point (marked with a star) $P_1: (T, \Omega_{\rm{m}},\Omega_{\rm{r}})=(0.449833,  0., 0.)$ (PEDE) and $P_1: (T, \Omega_{\rm{m}},\Omega_{\rm{r}})=(0.472999,  0., 0.)$ (GEDE).}
    \label{fig:3D}
\end{figure}

There are three solutions $P_3$, $P_4$ and $P_5$ dominated by radiation, dark matter and dark energy, respectively, that satisfy $T=1$. This means that $H=0$ at these solutions and they are saddles.  
\\
The point $P_6$ is the global source, it satisfies $\Omega_{\rm{m}}=0, \Omega_{\rm{r}}=0$, therefore, it is dominated by DE. As $T=0$, this implies that $H\rightarrow \infty$. Because it is a source, it represents the initial stages of the cosmic evolution, dominated by DE. This means that for the model not only dark energy accounts for the recent accelerated phase of the evolution  but also for the initial expanding phase.
\\
To analyse the the non-hyperbolic points $P_7$, $P_8$ and $P_9$ that satisfy $T\rightarrow 0$, we use numerical examination, where we have shown they are saddles (see Fig. \ref{fig:3D}). However, when the dynamics is restricted to the invariant set  $T=0$, it is governed by an integrable 2D dynamical system, such that the line $P_7: \left(0,\Omega_{\rm{m}},  1-\Omega_{\rm{m}}\right)$, along with the endpoints $P_8$ and $P_9$ are local attractors, whereas $P_6$ is the global source. The dynamics is exactly the same as presented in Fig. \ref{fig:PEDEphase-plot_T_0} after $\bar{\tau}$ is re-scaled by the factor $\frac{1-\Omega_{\rm{m}}^{(0)}-\Omega_{\rm{r}}^{(0)}}{\Delta
g(\Delta, z_t))}>0$.

Figure \ref{fig:3D} shows the dynamics of the systems \eqref{EQ.18} for the PEDE model with $\left(\Omega_{\rm{m}}^{(0)},\Omega_{\rm{r}}^{(0)}\right)=\left(0.252,7.62\times 10^{-5}\right)$ and \eqref{EQ.32} for the GEDE model with $\left(\Omega_{\rm{m}}^{(0)},\Omega_{\rm{r}}^{(0)}\right)=\left(0.247, 7.72\times 10^{-5}\right)$. The blue lines correspond to orbits with initial condition $\left(T(0),\Omega_{\rm{m}}{(0)},\Omega_{\rm{r}}{(0)}\right)=\left(0.5,0.252,7.62\times 10^{-5}\right)$ and $\left(0.5, 0.247, 7.72\times 10^{-5}\right)$, for PEDE and GEDE respectively, which represent the current universe.  All orbits are attracted by the point (marked with a star) $P_1: (T, \Omega_{\rm{m}},\Omega_{\rm{r}})=(0.449833,  0., 0.)$ (PEDE) and $P_1: (T, \Omega_{\rm{m}},\Omega_{\rm{r}})=(0.472999,  0., 0.)$ (GEDE). We have evaluated $g(\Delta, z_t)\sim 1.213046$ using the upper bounds of $z_t \sim 0.461$ and $\Delta \sim 1.314$ (homogeneous OHD). For GEDE, using DA OHD, we have the upper bounds $z_t= 0.277$ and $\Delta= 6.234$ given by \eqref{ZT}, \eqref{D} and  $\Omega_{DE}^{(0)} \approx 1-\Omega_{\rm{m}}^{(0)}=0.6801^{+0.039}_{-0.036}< 0.954722$. For the best-fit values of $z_t$ and $\Delta$, the interval is narrowed to $0\leq \Omega_{DE}^{(0)} < 0.789332$. Therefore, the attractor $P_2$ marginally exists. 

\begin{figure}
\includegraphics[width=0.485\textwidth]{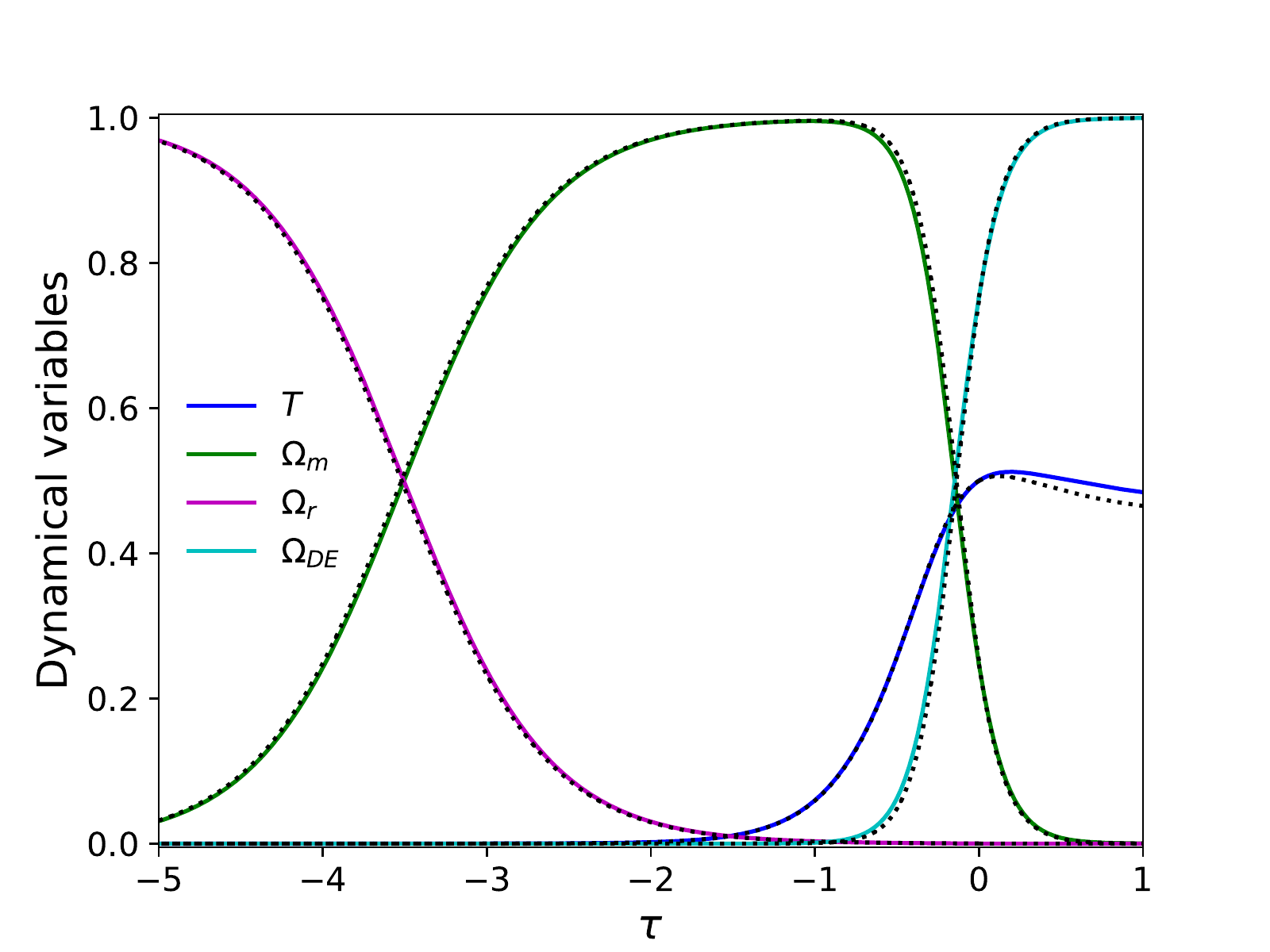} \\
\includegraphics[width=0.455\textwidth]{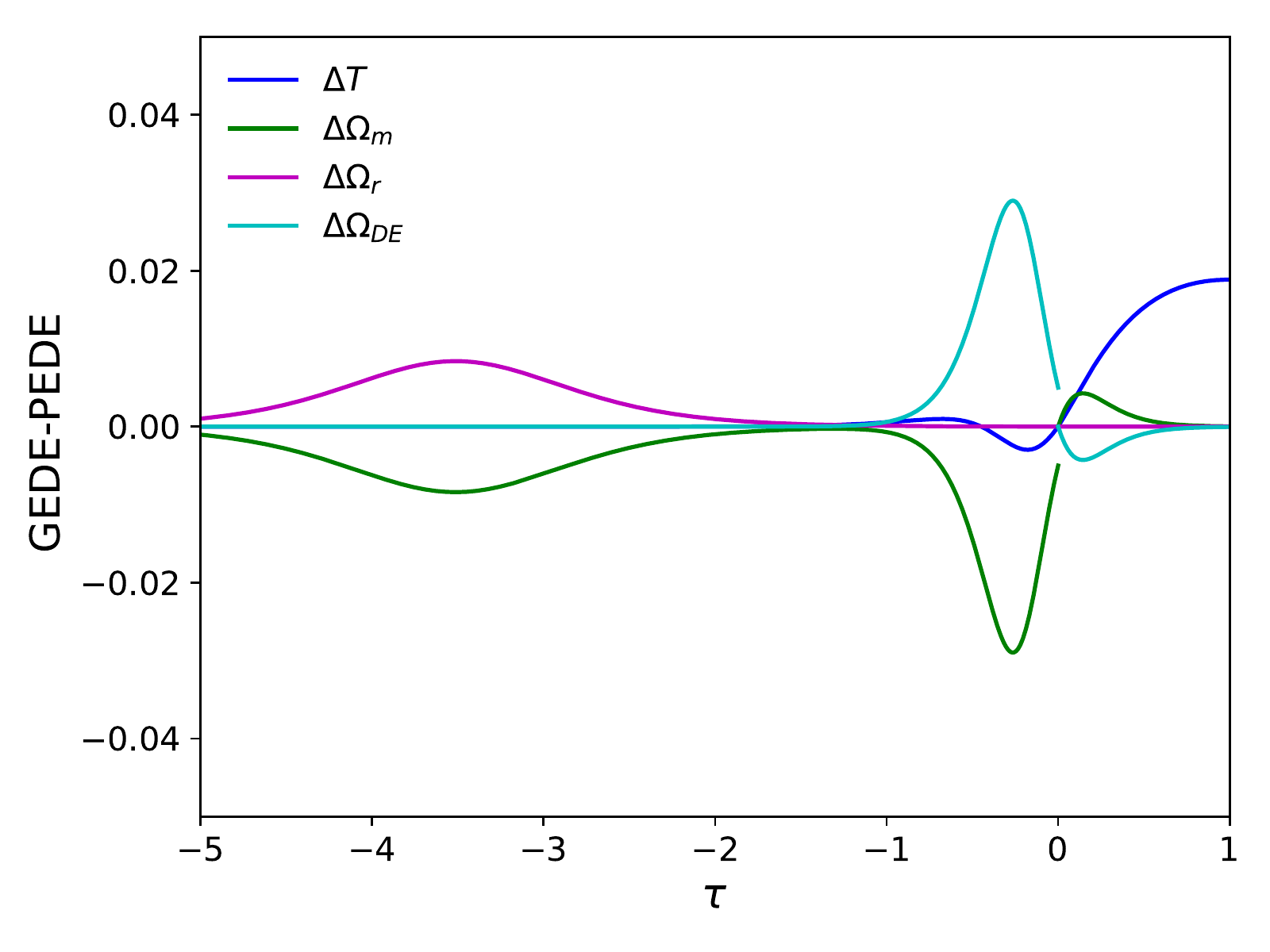}
\caption{Top panel. Evolution of the dynamical variables $(T, \Omega_m, \Omega_r, \Omega_{DE})$ over $\tau$ for the GEDE model. In dotted-black lines are the corresponding variables for the PEDE model. Bottom panel. $\Delta\Omega_i = \Omega^{GEDE}_i - \Omega^{PEDE}_i$ for $i=m, r, DE$ and $\Delta T = T^{GEDE}-T^{PEDE}$.}
\label{fig:DS}
\end{figure}

The top panel of Figure \ref{fig:DS} shows the numerical solution for the system  \eqref{EQ.18} (PEDE)  and  \eqref{EQ.32} (GEDE) using the initial conditions at current epoch. For this particular solution, at early epochs, the universe is dominated by radiation (equilibrium point $P_8$), later on, the matter becomes equal to radiation, then it begins to dominate (equilibrium point $P_9$). At late times, the emergent DE dominates the Universe dynamics in a de Sitter phase (equilibrium point $P_1$).
The aforementioned radiation dominated solution $P_8$ and the matter dominated solution $P_4$ do have $T=0$. This means that $H\rightarrow \infty$ at these solutions, as expected ($H \sim \frac{1}{2 t}$ for the usual radiation dominated solution and $H\sim \frac{2}{3 t}$ for the usual matter dominated solution). In the bottom panel of the same figure, it is shown the difference between the dynamical variables for the GEDE and PEDE models.

%%%%%%%%%%%%%%%%%%%%%%%%%%%%%%%%%%%%%%%%%%%%%%
\section{Conclusions}\label{sec:conclusions}
%%%%%%%%%%%%%%%%%%%%%%%%%%%%%%%%%%%%%%%%%%%%%%

We investigated the phenomenological models recently proposed by \citet{PEDE:2019ApJ,PEDE:2020} for which the dark energy is negligible at very early times of the Universe, dubbed PEDE and GEDE models. The main characteristic of these models is that they emerge at late times sourcing the accelerated expansion of the Universe through $\Omega_{DE}(z)\propto \mathrm{tanh}(z)$. While in PEDE model there is no extra degree of freedom as the standard model, the GEDE model introduces one free parameter ($\Delta$) which plays an important role to recover the $\Lambda$CDM and PEDE dynamics when $\Delta=0$ and $\Delta=1$, respectively. 

We put observational constraints for the PEDE and GEDE models through the most recent observational Hubble data samples: one including non-homogeneous OHD points from BAO and other sample where they are homogeneous.
Our analysis was performed with flat and Gaussian priors on the dimensionless Hubble parameter at the today $h$. 
Our constraints for the PEDE model are consistent with those obtained by \citet{PEDE:2019ApJ}. We also find consistent values for $h$ and $\Omega_m^{(0)}$, within $1.2\sigma$, with those reported by \cite{Pan:2019hac}. Nevertheless, our $\Delta$ limits (e.g. $0.69^{+0.624}_{-0.457}$) are consistent with PEDE model but in tension at $1\sigma$ with $\Delta=1.13\pm 0.28$ obtained by \citet{PEDE:2020} from Planck and $H_{0}$ (R19) measurements. Considering the uncertainties on $\Delta$, there is no strong support of GEDE over the $\Lambda$ model when OHD (low redshift) are employed.
In addition, we also reconstructed the cosmic evolution for the deceleration and jerk parameters in the PEDE and GEDE scenarios. For both models, the deceleration parameter undergoes a phase transition from a a decelerated expansion to an accelerated one (at $z\sim 0.78, 0.8$). By construction PEDE and GEDE are dynamical dark energy models, hence the jerk parameter deviates from one. Furthermore, our values for the deceleration-acceleration transition redshift and currrent values of the cosmographic parameters $q_0$ and $j_0$ are in agreement with those reported in the literature \citep{Garcia-Aspeitia:2018fvw,Haridasu:2018,Almada:2019,Almada:2020}. 
Regarding our stability analysis, we reconstructed the evolution of the dynamical variables $\Omega_m$, $\Omega_r$, and $\Omega_{DE}$ for PEDE and GEDE models using the homogeneous constraints since they are less unbiased due to any underlying cosmology (see \S \ref{sec:data}). 
We obtain that they have a very similar dynamics (Fig. \ref{fig:DS}).
We see that the Universe evolves to a de Sitter solution, corresponding to the equilibrium point $P_1$ with $a_{+}(t)= a_0 e^{ \tilde{\Lambda} t}$, (see \S \ref{sec:stability}) from a matter dominated phase, preceded by a radiation dominated epoch. However, the  main difference with the evolution of the $\Lambda$CDM model is that the global source (equilibrium point $P_6$) is dominated by DE.  This means that for the model not only dark energy accounts for the recent accelerated phase of the evolution but also the initial stages are driven by a DE dominated accelerated expanding phase.  This feature of PEDE/GEDE models is not mentioned by \citet{PEDE:2019ApJ,PEDE:2020}.
Furthermore, there is a possibility to have an attractor in $P_2$, with $a_{-}(t)= a_0 e^{-\tilde{\Lambda} t}$, which is not an expanding solution for $H_0>0$ at late times.  However, this solution is not supported by data \citep{Planck2018} because the condition $T\geq 0$ requires $0\leq \Omega_{\rm{DE}}^{(0)}<\frac{1}{2}$ (PEDE, homogeneous OHD), or $0\leq \Omega_{\rm{DE}}^{(0)}<\frac{g(\Delta, z_t)}{2}\sim 0.606518$ (GEDE, homogeneous OHD). In addition, our constraints (homogeneous OHD)  are $\Omega_{DE}^{(0)}=0.748^{+0.016}_{-0.015}$, $\Omega_{r}^{(0)}=(7.63^{+0.24}_{+0.23})\times 10^{-5}$ for PEDE and $\Omega_{DE}^{(0)}=0.753^{+0.018}_{-0.017}$, $\Omega_{r}^{(0)}=(7.72^{+0.26}_{+0.25})\times 10^{-5}$ for GEDE, which makes the condition of existence for $P_2$ hardest to be satisfied. However, using DA OHD, the intervals for GEDE are $0\leq \Omega_{\rm{DE}}^{(0)}<\frac{g(\Delta, z_t)}{2}\sim 0.954722$ where $z_t= 0.277$ and $\Delta= 6.234$, and $\Omega_{DE}^{(0)} \approx 1-\Omega_{\rm{m}}^{(0)}= 0.6801^{+0.039}_{-0.036}$. For the best-fit values of $z_t$ and $\Delta$, the interval is narrowed to $0\leq \Omega_{DE}^{(0)}< 0.789332$. Therefore,  $P_2$ is (marginally) allowed from these observations and it is an attractor, in contrast with previous cases.

On the other hand, many emergent DE models as those studied by \cite{Garcia-Aspeitia:2019yod} based on unimodular gravity, predict a birth of DE in the reionization epoch at $z\sim 17$, where an excess of photons has been detected by EDGES \citep{Bowman:2018yin} that could imply new physics beyond the standard scenario. In this vein, the PEDE (GEDE) model could also emerge at the same epoch,  being in agreement with the unimodular gravity. At $z\sim 17$, the PEDE density is $\rho_{DE} \sim 10\% \rho_{c}^{(0)}$ (i.e.  $\widetilde{\Omega}_{DE}\sim 0.1$).
Finally, the early accelerated phase, a possible connection to the reionization epoch together with other observational constraints, like those related with $H_0$ tension \citep{Pan:2019hac}, could be transcendental for PEDE and GEDE models and they should be further investigated.

%%%%%%%%%%%%%%%%%%%%%%%%%%%%%%%%%%%%%%%%%
\section*{Acknowledgments}
We thank the anonymous referee for thoughtful remarks and suggestions. G.L.  was funded by ANID
through FONDECYT Iniciación grant no. 11180126 and by Vicerrectoría de Investigación y Desarrollo Tecnológico at
Universidad Católica del Norte., A.H.A. thanks to the PRODEP project, Mexico for resources and financial support. J.M. acknowledges the support from CONICYT project Basal AFB-170002, M.A.G.-A. acknowledges support from SNI-M\'exico, CONACyT research fellow, COZCyT and Instituto Avanzado de Cosmolog\'ia (IAC) collaborations.  V.M. acknowledges the support of Centro de Astrof\'{\i}sica de Valpara\'{\i}so (CAV). J.M., M.A.G.-A and V.M. acknowledge CONICYT REDES (190147). \\

\section*{Note added}
{While this work was being typed, we became aware of a complementary study of PEDE model, developed by \cite{Liu:2020vgn}, that appeared in the arXiv repository. \cite{Liu:2020vgn}, used CMB data from Planck 2018, BAO measurements and SNIa data, to obtain the bounds on total
neutrino masses with the approximation of degenerate neutrino masses, in some Dark Energy settings, in particular in PEDE models.}

\appendix
\section{PEDE model including the transition redshift $z_{t}$}
\label{sec:AppendixA}
\begin{figure}
\centering
\includegraphics[width=7cm,scale=0.5]{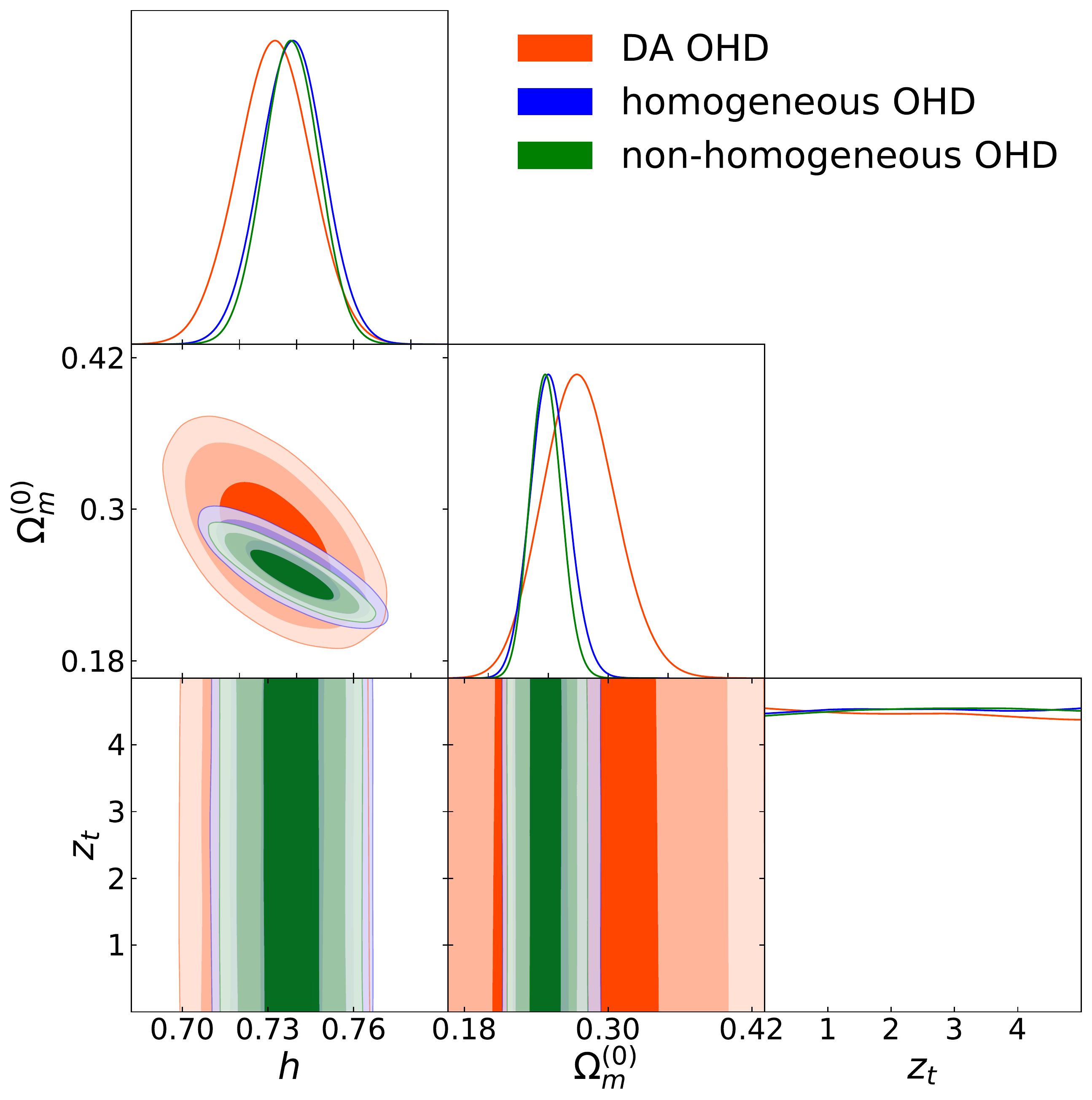}
\caption{1D posterior distributions and 2D contours of the free parameters for PEDE including $z_{t}$ as free parameter at $1\sigma$, $2\sigma$, $3\sigma$ CL (from darker to lighter respectively)} 
\label{fig:contours:PEDEzt}
\end{figure}

\cite{PEDE:2019ApJ} also introduced a transition redshift $z_{t}$ into $\widetilde{\Omega}_{\rm{DE}}(z)$ of PEDE as
\begin{equation}
\widetilde{\Omega}_{\rm{DE}}(z)\,=\,\Omega_{\rm{DE}}^{(0)} \frac{ 1 - {\rm{tanh}}\left( { \log}_{10}(\frac{1+z}{1+z_{t}}) \right)}{ 1 + {\rm{tanh}}\left({\log}_{10}(1+z_{t}) \right)},
\label{eq:omega_pede_with_zt}
\end{equation}
which satisfies $\widetilde{\Omega}_{DE}(z_t)=\Omega^{(0)}_{m}(1+z_t)^3$.
To assess the impact of this parameter $z_{t}$ in our PEDE constraints, we carry out the MCMC analysis for all the OHD samples using the same Gaussian prior on $h$ as before and including $z_{t}$ as free parameter with the flat prior: $\left[0,5\right]$. Figure \ref{fig:contours:PEDEzt} shows the 1D posterior distributions and 2D contours for $h$, $\Omega_{m}^{(0)}$ and $z_{t}$. Notice that the ($h$, $\Omega_m^{(0)}$) bounds presented are independent of the selected value for $z_t$. This same result was found by \citet{PEDE:2019ApJ}. 

\bibliographystyle{mnras}
\bibliography{main}

\end{document}